\documentclass[graybox]{svmult}
\usepackage{graphicx}
\usepackage{feynmp-auto}

 \usepackage{hyperref}
\hypersetup{colorlinks=true,linkcolor=magenta,anchorcolor=green,citecolor=cyan,filecolor=black,menucolor=black,urlcolor=black}

\usepackage{amsmath}
\usepackage{amscd}
\usepackage{amssymb}
\usepackage{bm}
\usepackage{xcolor}
\usepackage{wasysym}
\usepackage{url}
\usepackage{amsfonts}
\usepackage[english]{babel}
\usepackage[T1]{fontenc}
\usepackage[latin1]{inputenc}
\usepackage{ae}

\usepackage{url}

\usepackage{color}
\usepackage{cite}
\usepackage{mathptmx}       % selects Times Roman as basic font
\usepackage{helvet}         % selects Helvetica as sans-serif font
\usepackage{courier}        % selects Courier as typewriter font
\usepackage{type1cm}        % activate if the above 3 fonts are
\usepackage{makeidx}         % allows index generation
\usepackage{graphicx}        % standard LaTeX graphics tool
\usepackage{multicol}        % used for the two-column index
\usepackage[bottom]{footmisc}% places footnotes at page bottom
\newcommand{\Li}{{\rm Li}}

    \usepackage{upquote} % Upright quotes for verbatim code
 \usepackage{fancyvrb} % verbatim replacement that allows latex
  
    \DefineVerbatimEnvironment{Highlighting}{Verbatim}{commandchars=\\\{\}}
    \definecolor{incolor}{rgb}{0.0, 0.0, 0.5}
    \definecolor{outcolor}{rgb}{0.545, 0.0, 0.0}

\makeatletter
\def\PY@reset{\let\PY@it=\relax \let\PY@bf=\relax%
    \let\PY@ul=\relax \let\PY@tc=\relax%
    \let\PY@bc=\relax \let\PY@ff=\relax}
\def\PY@tok#1{\csname PY@tok@#1\endcsname}
\def\PY@toks#1+{\ifx\relax#1\empty\else%
    \PY@tok{#1}\expandafter\PY@toks\fi}
\def\PY@do#1{\PY@bc{\PY@tc{\PY@ul{%
    \PY@it{\PY@bf{\PY@ff{#1}}}}}}}
\def\PY#1#2{\PY@reset\PY@toks#1+\relax+\PY@do{#2}}

\expandafter\def\csname PY@tok@w\endcsname{\def\PY@tc##1{\textcolor[rgb]{0.73,0.73,0.73}{##1}}}
\expandafter\def\csname PY@tok@c\endcsname{\let\PY@it=\textit\def\PY@tc##1{\textcolor[rgb]{0.25,0.50,0.50}{##1}}}
\expandafter\def\csname PY@tok@cp\endcsname{\def\PY@tc##1{\textcolor[rgb]{0.74,0.48,0.00}{##1}}}
\expandafter\def\csname PY@tok@k\endcsname{\let\PY@bf=\textbf\def\PY@tc##1{\textcolor[rgb]{0.00,0.50,0.00}{##1}}}
\expandafter\def\csname PY@tok@kp\endcsname{\def\PY@tc##1{\textcolor[rgb]{0.00,0.50,0.00}{##1}}}
\expandafter\def\csname PY@tok@kt\endcsname{\def\PY@tc##1{\textcolor[rgb]{0.69,0.00,0.25}{##1}}}
\expandafter\def\csname PY@tok@o\endcsname{\def\PY@tc##1{\textcolor[rgb]{0.40,0.40,0.40}{##1}}}
\expandafter\def\csname PY@tok@ow\endcsname{\let\PY@bf=\textbf\def\PY@tc##1{\textcolor[rgb]{0.67,0.13,1.00}{##1}}}
\expandafter\def\csname PY@tok@nb\endcsname{\def\PY@tc##1{\textcolor[rgb]{0.00,0.50,0.00}{##1}}}
\expandafter\def\csname PY@tok@nf\endcsname{\def\PY@tc##1{\textcolor[rgb]{0.00,0.00,1.00}{##1}}}
\expandafter\def\csname PY@tok@nc\endcsname{\let\PY@bf=\textbf\def\PY@tc##1{\textcolor[rgb]{0.00,0.00,1.00}{##1}}}
\expandafter\def\csname PY@tok@nn\endcsname{\let\PY@bf=\textbf\def\PY@tc##1{\textcolor[rgb]{0.00,0.00,1.00}{##1}}}
\expandafter\def\csname PY@tok@ne\endcsname{\let\PY@bf=\textbf\def\PY@tc##1{\textcolor[rgb]{0.82,0.25,0.23}{##1}}}
\expandafter\def\csname PY@tok@nv\endcsname{\def\PY@tc##1{\textcolor[rgb]{0.10,0.09,0.49}{##1}}}
\expandafter\def\csname PY@tok@no\endcsname{\def\PY@tc##1{\textcolor[rgb]{0.53,0.00,0.00}{##1}}}
\expandafter\def\csname PY@tok@nl\endcsname{\def\PY@tc##1{\textcolor[rgb]{0.63,0.63,0.00}{##1}}}
\expandafter\def\csname PY@tok@ni\endcsname{\let\PY@bf=\textbf\def\PY@tc##1{\textcolor[rgb]{0.60,0.60,0.60}{##1}}}
\expandafter\def\csname PY@tok@na\endcsname{\def\PY@tc##1{\textcolor[rgb]{0.49,0.56,0.16}{##1}}}
\expandafter\def\csname PY@tok@nt\endcsname{\let\PY@bf=\textbf\def\PY@tc##1{\textcolor[rgb]{0.00,0.50,0.00}{##1}}}
\expandafter\def\csname PY@tok@nd\endcsname{\def\PY@tc##1{\textcolor[rgb]{0.67,0.13,1.00}{##1}}}
\expandafter\def\csname PY@tok@s\endcsname{\def\PY@tc##1{\textcolor[rgb]{0.73,0.13,0.13}{##1}}}
\expandafter\def\csname PY@tok@sd\endcsname{\let\PY@it=\textit\def\PY@tc##1{\textcolor[rgb]{0.73,0.13,0.13}{##1}}}
\expandafter\def\csname PY@tok@si\endcsname{\let\PY@bf=\textbf\def\PY@tc##1{\textcolor[rgb]{0.73,0.40,0.53}{##1}}}
\expandafter\def\csname PY@tok@se\endcsname{\let\PY@bf=\textbf\def\PY@tc##1{\textcolor[rgb]{0.73,0.40,0.13}{##1}}}
\expandafter\def\csname PY@tok@sr\endcsname{\def\PY@tc##1{\textcolor[rgb]{0.73,0.40,0.53}{##1}}}
\expandafter\def\csname PY@tok@ss\endcsname{\def\PY@tc##1{\textcolor[rgb]{0.10,0.09,0.49}{##1}}}
\expandafter\def\csname PY@tok@sx\endcsname{\def\PY@tc##1{\textcolor[rgb]{0.00,0.50,0.00}{##1}}}
\expandafter\def\csname PY@tok@m\endcsname{\def\PY@tc##1{\textcolor[rgb]{0.40,0.40,0.40}{##1}}}
\expandafter\def\csname PY@tok@gh\endcsname{\let\PY@bf=\textbf\def\PY@tc##1{\textcolor[rgb]{0.00,0.00,0.50}{##1}}}
\expandafter\def\csname PY@tok@gu\endcsname{\let\PY@bf=\textbf\def\PY@tc##1{\textcolor[rgb]{0.50,0.00,0.50}{##1}}}
\expandafter\def\csname PY@tok@gd\endcsname{\def\PY@tc##1{\textcolor[rgb]{0.63,0.00,0.00}{##1}}}
\expandafter\def\csname PY@tok@gi\endcsname{\def\PY@tc##1{\textcolor[rgb]{0.00,0.63,0.00}{##1}}}
\expandafter\def\csname PY@tok@gr\endcsname{\def\PY@tc##1{\textcolor[rgb]{1.00,0.00,0.00}{##1}}}
\expandafter\def\csname PY@tok@ge\endcsname{\let\PY@it=\textit}
\expandafter\def\csname PY@tok@gs\endcsname{\let\PY@bf=\textbf}
\expandafter\def\csname PY@tok@gp\endcsname{\let\PY@bf=\textbf\def\PY@tc##1{\textcolor[rgb]{0.00,0.00,0.50}{##1}}}
\expandafter\def\csname PY@tok@go\endcsname{\def\PY@tc##1{\textcolor[rgb]{0.53,0.53,0.53}{##1}}}
\expandafter\def\csname PY@tok@gt\endcsname{\def\PY@tc##1{\textcolor[rgb]{0.00,0.27,0.87}{##1}}}
\expandafter\def\csname PY@tok@err\endcsname{\def\PY@bc##1{\setlength{\fboxsep}{0pt}\fcolorbox[rgb]{1.00,0.00,0.00}{1,1,1}{\strut ##1}}}
\expandafter\def\csname PY@tok@kc\endcsname{\let\PY@bf=\textbf\def\PY@tc##1{\textcolor[rgb]{0.00,0.50,0.00}{##1}}}
\expandafter\def\csname PY@tok@kd\endcsname{\let\PY@bf=\textbf\def\PY@tc##1{\textcolor[rgb]{0.00,0.50,0.00}{##1}}}
\expandafter\def\csname PY@tok@kn\endcsname{\let\PY@bf=\textbf\def\PY@tc##1{\textcolor[rgb]{0.00,0.50,0.00}{##1}}}
\expandafter\def\csname PY@tok@kr\endcsname{\let\PY@bf=\textbf\def\PY@tc##1{\textcolor[rgb]{0.00,0.50,0.00}{##1}}}
\expandafter\def\csname PY@tok@bp\endcsname{\def\PY@tc##1{\textcolor[rgb]{0.00,0.50,0.00}{##1}}}
\expandafter\def\csname PY@tok@fm\endcsname{\def\PY@tc##1{\textcolor[rgb]{0.00,0.00,1.00}{##1}}}
\expandafter\def\csname PY@tok@vc\endcsname{\def\PY@tc##1{\textcolor[rgb]{0.10,0.09,0.49}{##1}}}
\expandafter\def\csname PY@tok@vg\endcsname{\def\PY@tc##1{\textcolor[rgb]{0.10,0.09,0.49}{##1}}}
\expandafter\def\csname PY@tok@vi\endcsname{\def\PY@tc##1{\textcolor[rgb]{0.10,0.09,0.49}{##1}}}
\expandafter\def\csname PY@tok@vm\endcsname{\def\PY@tc##1{\textcolor[rgb]{0.10,0.09,0.49}{##1}}}
\expandafter\def\csname PY@tok@sa\endcsname{\def\PY@tc##1{\textcolor[rgb]{0.73,0.13,0.13}{##1}}}
\expandafter\def\csname PY@tok@sb\endcsname{\def\PY@tc##1{\textcolor[rgb]{0.73,0.13,0.13}{##1}}}
\expandafter\def\csname PY@tok@sc\endcsname{\def\PY@tc##1{\textcolor[rgb]{0.73,0.13,0.13}{##1}}}
\expandafter\def\csname PY@tok@dl\endcsname{\def\PY@tc##1{\textcolor[rgb]{0.73,0.13,0.13}{##1}}}
\expandafter\def\csname PY@tok@s2\endcsname{\def\PY@tc##1{\textcolor[rgb]{0.73,0.13,0.13}{##1}}}
\expandafter\def\csname PY@tok@sh\endcsname{\def\PY@tc##1{\textcolor[rgb]{0.73,0.13,0.13}{##1}}}
\expandafter\def\csname PY@tok@s1\endcsname{\def\PY@tc##1{\textcolor[rgb]{0.73,0.13,0.13}{##1}}}
\expandafter\def\csname PY@tok@mb\endcsname{\def\PY@tc##1{\textcolor[rgb]{0.40,0.40,0.40}{##1}}}
\expandafter\def\csname PY@tok@mf\endcsname{\def\PY@tc##1{\textcolor[rgb]{0.40,0.40,0.40}{##1}}}
\expandafter\def\csname PY@tok@mh\endcsname{\def\PY@tc##1{\textcolor[rgb]{0.40,0.40,0.40}{##1}}}
\expandafter\def\csname PY@tok@mi\endcsname{\def\PY@tc##1{\textcolor[rgb]{0.40,0.40,0.40}{##1}}}
\expandafter\def\csname PY@tok@il\endcsname{\def\PY@tc##1{\textcolor[rgb]{0.40,0.40,0.40}{##1}}}
\expandafter\def\csname PY@tok@mo\endcsname{\def\PY@tc##1{\textcolor[rgb]{0.40,0.40,0.40}{##1}}}
\expandafter\def\csname PY@tok@ch\endcsname{\let\PY@it=\textit\def\PY@tc##1{\textcolor[rgb]{0.25,0.50,0.50}{##1}}}
\expandafter\def\csname PY@tok@cm\endcsname{\let\PY@it=\textit\def\PY@tc##1{\textcolor[rgb]{0.25,0.50,0.50}{##1}}}
\expandafter\def\csname PY@tok@cpf\endcsname{\let\PY@it=\textit\def\PY@tc##1{\textcolor[rgb]{0.25,0.50,0.50}{##1}}}
\expandafter\def\csname PY@tok@c1\endcsname{\let\PY@it=\textit\def\PY@tc##1{\textcolor[rgb]{0.25,0.50,0.50}{##1}}}
\expandafter\def\csname PY@tok@cs\endcsname{\let\PY@it=\textit\def\PY@tc##1{\textcolor[rgb]{0.25,0.50,0.50}{##1}}}

\def\PYZca{\char`\^}

\def\PYZhy{\char`\-}

\makeatother

\newcount\hh
\newcount\mm
\mm=\time
\hh=\time
\divide\hh by 60
\divide\mm by 60
\multiply\mm by 60
\mm=-\mm
\advance\mm by \time
\def\thistime{\number\hh:\ifnum\mm<10{}0\fi\number\mm}
\def\eqref#1{(\ref{#1})}
\def\su{\circleddash}

\def\Li#1(#2){\textrm{Li}_{#1}\left(#2\right)}
\def\cLi_#1(#2){\mathcal{L}_{#1}\left(#2\right)}
\def\bLi_#1(#2){\mathbf{L}_{#1}\left(#2\right)}

\def\cEs{\mathcal E_\circleddash}

\def\IR{{\mathbb R}}

\def\IN{{\mathbb N}}
\def\IP{{\mathbb P}}

\def\cU{\mathcal{U}}

\def\cE{\mathcal{E}}

\def\cF{\mathcal{F}}

\def\Imm{\Im\textrm{m}}

\makeindex             % used for the subject index
\begin{document}

\title*{Feynman integrals, toric geometry and mirror symmetry}
\author{Pierre Vanhove\thanks{IPHT-t18/096}}
 \institute{ CEA, DSM, Institut de Physique Th{\'e}orique, IPhT, CNRS, MPPU,
URA2306, Saclay, F-91191 Gif-sur-Yvette, France \\
National Research University Higher School of Economics, Russian Federation}
\maketitle

%%%%%%%%%%%%%%%%%%%%%%%%%%%%%%%%%%%%%%%%%%%%%%%%%%%%%%%%%%%%%%%%%
\abstract{
% {{\bf Draft version \today\ at \thistime}
This expository
  text is about using toric geometry and mirror symmetry for
  evaluating Feynman integrals. We show that the maximal cut of a
  Feynman integral is a GKZ hypergeometric series.  We explain how
  this allows to   determine the minimal differential operator acting on the Feynman
  integrals.  We illustrate the method on sunset integrals in
  two dimensions at various loop orders. The graph polynomials of the multi-loop sunset Feynman graphs lead to reflexive
  polytopes containing the origin and the associated variety are
  ambient spaces for Calabi-Yau hypersurfaces. Therefore the sunset family is a natural home
  for mirror symmetry techniques.   We review the
  evaluation of the two-loop sunset integral as an elliptic
  dilogarithm and as a trilogarithm. The equivalence between these two
  expressions is a consequence of 1)  the local mirror symmetry for the
  non-compact Calabi-Yau three-fold obtained as
  the anti-canonical hypersurface
of the del Pezzo surface of
  degree 6 defined by the sunset graph polynomial  and  2) that the sunset Feynman integral is expressed in terms of
  the local Gromov-Witten prepotential of this del Pezzo surface.  }
%%%%%%%%%%%%%%%%%%%%%%%%%%%%%%%%%%%%%%%%%%%%%%%%%%%%%%%%%%%%%%%%%

%
%  --> relevant terms, to appear in the index shall alwys be put into \index{....}
%

%-------------------------------------------------------------------------
\section{Introduction}

Scattering amplitudes are fundamental quantities used to understand
fundamental interactions and the elementary constituents in
Nature.  It is well known that  scattering amplitudes are used in particle physics to compare
the theoretical predictions to experimental measurements in particle
colliders (see~\cite{Bendavid:2018nar} for instance).  More
recently the use of modern developments in  scattering amplitudes  have
been extended to gravitational physics like unitarity methods  to gravitational wave physics~\cite{Neill:2013wsa,Bjerrum-Bohr:2013bxa,Cachazo:2017jef,Guevara:2017csg,Bjerrum-Bohr:2018xdl}.

The $l$-loop scattering amplitude $A_{n, l}^D(\underline s,\underline m^2 )$ between $n$ fields in $D$
dimensions is a function of the kinematics invariants
$\underline s=\{s_{ij}=(p_i+p_j)^2,  1\leq i,j\leq n\}$ where $p_i$ are the
incoming momenta of the external particle, and the   internal masses $\underline
m^2=(m_1,\dots,m_r)$.

\medskip
We focus on the questions :  what kind of function is a
Feynman integral?  What is the best analytic representation?

\medskip

The answer to these questions depend very strongly on which properties
one wants to display.  An analytic representation suitable for an high precision
numerical evaluation may not be the one that displays the mathematical
nature of the integral.

For  instance the two-loop sunset integral  has received many different,
but equivalent, analytical expressions:
hypergeometric and Lauricella functions~\cite{Tarasov:2006nk,Bauberger:1994by}, Bessel integral representation~\cite{Bailey:2008ib, Broadhurst:2008mx,Broadhurst:2016myo}, Elliptic
integrals~\cite{Caffo:1999nk,Laporta:2004rb}, Elliptic
polylogarithms~\cite{Adams:2014vja,Adams:2015gva,Adams:2015ydq,Adams:2016vdo,Adams:2018ulb,Adams:2018yqc,Bloch:2016izu}
and trilogarithms~\cite{Bloch:2016izu}.

The approach that we will follow here will be guided by the geometry
of the graph polynomial using the parametric representation of the
Feynman integral.
In \S\ref{sec:feynman-integral} we review the description of the
Feynman integral $I_\Gamma$ for a graph $\Gamma$ in parametric space. We focus on the properties of
the second Symanzik polynomial as a preparation for the toric approach
used in \S\ref{sec:toric}. In \S\ref{sec:Maximal cut} we
show that the maximal cut  $\pi_\Gamma$ of a Feynman integral has a parametric
representation similar to the one of the Feynman integral $I_\Gamma$
where the only difference is the cycle of
integration. The toric geometry approach is described in~\S\ref{sec:toric}. In
section~\ref{sec:gzk-method} we explain that the maximal cut integral
is an hypergeometric series from the
Gel'fand-Kapranov-Zelevinski (GKZ) construction. In
section~\ref{sec:diff-oper}, we show on examples how to derive the
minimal  differential operator annihilating the maximal cut
integral. In \S\ref{sec:sunset}  we review the evaluation of the
two-loop sunset integral in two space-time dimensions. In
section~\ref{sec:sunsetdilog} we give its expression as an elliptic
dilogarithm
\begin{equation}
  I_\su(p^2,\xi_1^2,\xi_2^2,\xi_3^2) \propto \varpi\, \sum_{i=1}^6 c_i \sum_{n\geq1}
  (\textrm{Li}_2(q^n z_i)- (\textrm{Li}_2(-q^n z_i))
\end{equation}
where $\varpi$ is a period of the elliptic curve defined by the graph
polynomial, $q$ the nome function of the external momentum $p^2$ and internal
masses $\xi_i^2$ for $i=1,2,3$.
In \S\ref{sec:sunsettrilog} we show that the sunset integral
evaluates as sum of trilogarithm functions in~\eqref{e:3triexp}
\begin{multline}
  I_\su(p^2,\xi_1^2,\xi_2^2,\xi_3^2) \propto \varpi\,(3(\log Q)^3\cr
  + \sum_{(n_1,n_2,n_3)\geq0}
     \left(d_{n_1,n_2,n_3}+\delta_{n_1,n_2,n_3}\log(-p^2)\right)
     \textrm{Li}_3(\prod_{i=1}^3 \xi_i^{2n_i}Q^{n_i})\,.
\end{multline}
 In \S\ref{sec:mirror} we show that the equivalence
between theses two expression is the result of a local mirror map,
$q\leftrightarrow Q$ in~\eqref{e:mirrormap}, for
the non-compact Calabi-Yau three-fold obtained as the anti-canonical
bundle over the del Pezzo 6 surface defined by the sunset graph
polynomial. Remarkably the sunset Feynman integral  is expressed in terms of the  genus
zero local
Gromov-Witten prepotential~\cite{Bloch:2016izu}.
Therefore this provides a natural application
for Batyrev's mirror symmetry
techniques~\cite{Batyrev94}.
One remarkable fact is that the computation can be done
using the existing technology of mirror symmetry developed in other
physical~\cite{Hosono:1993qy,CKYZ,Huang:2013yta} or
mathematics~\cite{Doran:2011ti} contexts.

%-------------------------------------------------------------------------
\section{{\bf Feynman integrals}}
\label{sec:feynman-integral}

A connected Feynman graph $\Gamma$ is determined by the number  $n$ of propagators
(internal edges),
the number $l$ of  loops, and the number $v$ of
vertices. The Euler characteristic of the graph relates these three
numbers as $v=n-l+1$, therefore only the number of loops $l$ and the
number $n$ of propagators are needed.

In a momentum representation an $l$-loop with $n$ propagators Feynman
graph reads
\begin{equation}\label{e:GraphFeyn}
  I_\Gamma(\underline s,\underline \xi^2,\underline \nu,D):=
 { (\mu^2)^{\omega}\over \pi^{lD\over2}}\,{\prod_{i=1}^n
    \Gamma(\nu_i)\over \Gamma(\omega)} \,
  \int_{(\IR^{1,D-1})^l} \,   {\prod_{i=1}^l  d^D\ell_i\over
    \prod_{i=1}^n (q_i^2-m_i^2+i\varepsilon)^{\nu_i}}\,,
\end{equation}
where $D$ is the space-time dimension, and we set
$\omega:=\sum_{i=1}^n \nu_i-lD/2$ and $q_i$ is the momentum flowing
in between the vertices $i$ and $i+1$.   With $\mu^2$  a scale of dimension
mass squared. From now we set $m_i^2=\xi_i^2\mu^2$ and
$p_i\to p_{i}\mu$, with these new variables the $\mu^2$ dependence
disappear.  The internal masses are positive $\xi_i^2\geq0$ with
$1\leq i\leq n$. Finally $+i\varepsilon$ with $\varepsilon>0$ is the
Feynman prescription for the propagators for a space-time metric of
signature $(+-\cdots-)$. The arguments of the Feynman integral are
$\underline \xi^2:=\{\xi_1^2,\dots,\xi_n^2\}$ and
$\nu:=\{\nu_1,\dots,\nu_n\}$ and
$\underline s:=\{s_{ij}=(p_i+p_j)^2\}$ with $p_i$ with $i=1,\dots,v_e$
with $0\leq v_e\leq v$ the external momenta subject to the momentum
conservation condition $p_1+\cdots+p_{v_e}=0$.
There are $n$ internal masses $\xi_i^2$ with $1\leq i\leq n$, is
$v_e$ is the number of external momenta we have $v_e$ external masses
$p_i^2$ with $1\leq i\leq v_e$  (some of the mass could vanish but we
do a generic counting here),   and ${v_e(v_e-3)\over 2}$ independent
kinematics invariants $s_{ij}=(p_i+p_j)^2$.
The total number of kinematic parameters is
\begin{equation}
 N_\Gamma(n,l)= n+{v_e(v_e-1)\over2}\leq N_\Gamma(n,l)^{max}= n+{(n-l+1)(n-l)\over2}  \,.
\end{equation}
We set
\begin{equation}
 I_\Gamma(\underline s,\underline m,D):= I_\Gamma(\underline
 s,\underline m,1,\dots,1,D) \, ,
\end{equation}
and for  $\nu_i$ positive integers we have
\begin{equation}
    I_\Gamma(\underline s,\underline m,\underline \nu,D)=
    \prod_{i=1}^n \left({\partial\over \partial (\xi_i^2)}\right)^{\nu_i} I_\Gamma(\underline s,\underline m,D)\,.
\end{equation}

\subsection{The parametric representation}
\label{sec:feynm-param}

Introducing the variables $x_i$ with $1\leq i\leq n$ such that
\begin{equation}
  \sum_{i=1}^n x_i (q_i^2-\xi_i^2)=
  (\ell^\mu_1,\dots,\ell^\mu_l)\cdot \Omega\cdot
  (\ell^\mu_1,\dots,\ell^\mu_l)^T\!+
  (\ell^\mu_1,\dots,\ell^\mu_l)\cdot (Q_1^\mu,\dots,Q_l^\mu)-J,
\end{equation}
and performing standard Gaussian integrals on the $x_i$ (see~\cite{Vanhove:2014wqa} for
instance) one obtains the equivalent parametric representation that we
will use in these notes

\begin{equation}\label{e:GraphPara2}
    I_\Gamma(\underline s,\underline \xi,\underline \nu,D)= \int_{\Delta_n}  \Omega_\Gamma,
\end{equation}
the integrand is the $n-1$-form
\begin{equation}
  \Omega_\Gamma=   \prod_{i=1}^n x_i^{\nu_i-1} \,
  {\cU^{\omega- {D\over2}}\over \cF^{\omega}}\, \Omega_0,
\end{equation}
where $\Omega_0$ is the differential $n-1$-form on the real projective space $\mathbb P^{n-1}$

\begin{equation}\label{e:diffInt}
  \Omega_0:= \sum_{j=1}^n  (-1)^{j-1} \, x_j\, dx_1\wedge \cdots \wedge
  \widehat {dx_j}\wedge\cdots \wedge dx_n  \,,
\end{equation}
where $\widehat{dx_j}$ means that $dx_j$ is omitting in this sum.
The domain of integration $\Delta_n$ is defined as
\begin{equation}\label{e:DefDomain}
  \Delta_n:=\{[x_1,\cdots,x_n]\in \mathbb P^{n-1}| x_i\in\mathbb R,  x_i\geq0\}.
\end{equation}
The second Symanzik polynomial
$\mathcal F=\mathcal U\,
\bigl((Q_1^\mu,\dots,Q_l^\mu)\cdot \Omega^{-1}\cdot
  (Q_1^\mu,\dots,Q_l^\mu)^T\\-J\bigr)$,
takes the form
\begin{equation}\label{e:Fdef}
  \cF(\underline s,\underline \xi^2,x_1,\dots,x_n)= \mathcal U(x_1,\dots,x_n)\, \left(\sum_{i=1}^n \xi_i^2 x_i\right) - \sum_{1\leq
    i\leq j\leq n} s_{ij} \mathcal G_{ij}(x_1,\dots,x_n)
\end{equation}
where the first Symanzik polynomial $\mathcal U(x_1,\dots,x_n)=\det\Omega$ and
$\mathcal G_{ij}(x_1,\dots,x_n)$ are polynomial in the $x_i$ variables only.

\medskip

\begin{itemize}

\item The first Symanzik polynomial $\mathcal U(x_1,\dots,x_n)$ is an
  homogeneous polynomial of degree $l$ in the Feynman parameters $x_i$
  and it is at most linear in each of the $x_i$ variables. It does not
  depend on the physical parameters. This polynomial is also known as
  the Kirchhoff polynomial of graph $\Gamma$. Which is as well the
  determinant of the Laplacian of the graph
  see~\cite[eq~(35)]{Bogner:2010kv} for a definition.
\item The polynomial $\mathcal U(x_1,\dots,x_n)$ can be seen as the
  determinant of the period matrix $\Omega$ of the punctured Feynman graph~\cite{Vanhove:2014wqa},
  i.e. the graph with amputated external legs. Or equivalently
  it can be obtained by considering the degeneration limit of a genus
  $l$ Riemann surfaces with $n$ punctures. This connection plays an
  important in understanding the quantum field theory Feynman
  integrals as the $\alpha'\to0$ limit of the corresponding string
  theory integrals~\cite{Tourkine:2013rda,Amini:2015czm}.

\item The graph polynomial $\cF$ is homogeneous of degree $l+1$ in the
  variables $(x_1,\dots,x_n)$. This polynomial
  depends on the internal masses $\xi_i^2$ and the kinematic
  invariants $s_{ij}=(p_i\cdot p_j)/\mu^2$. The polynomials $\mathcal
  G_{ij}$ are at most  linear in all
  the variables $x_i$ since this is given by the spanning 2-trees~\cite{Bogner:2010kv}. Therefore  if all internal masses are vanishing then
  $\cF$ is linear in the Feynman parameters $x_i$.

\item The $\cU$ and $\cF$ are independent of the dimension of
space-time. The space-time dimension enters only in the powers of
$\cU$ and $\cF$ in the parametric representation for the Feynman
graphs. Therefore one can see the Feynman integral as a meromorphic
function of $(\underline \nu,D)$  in $\mathbb C^{1+n}$ as discussed in~\cite{Speer}.

\item All the physical parameters, the internal masses $\xi_i^2$ and
  the kinematic variables $s_{ij}=(p_i\cdot p_j)/\mu^2$ (that includes the
  external masses) enter linearly. This will be important for the
  toric approach described in \S\ref{sec:toric}.

\end{itemize}

%-------------------------------------------------------------------------
\subsection{{\bf Maximal cut}}\label{sec:Maximal cut}

We show that the maximal cut of a Feynman graph has a nice parametric
representation.
Let us consider the maximal cut
\begin{equation}\label{e:GraphFeynCut}
 \pi_\Gamma(\underline s,\underline \xi^2,D):=
 {1\over \Gamma(\omega)(2i\pi)^n\pi^{lD\over2}} \,
  \int_{(\IR^{1,D-1})^L} \,   \prod_{i=1}^l  d^D\ell_i
    \prod_{i=1}^n \delta(q_i^2-m_i^2+i\varepsilon),
\end{equation}
of the Feynman integral $I_\Gamma(\underline s,\underline \xi^2,D)$
which is obtained from the Feynman integral in~\eqref{e:GraphFeyn} by
replacing all propagators by a delta-function
\begin{align}
  {1\over d^2} ={1\over 2i\pi} \delta(d^2)  .
\end{align}
Using the representation of the $\delta$-function
\begin{equation}
  \delta(x)= \int_{-\infty}^{+\infty} dw e^{iwx}  ,
\end{equation}
we obtain that the integral is
\begin{equation}
  \pi_\Gamma(\underline s,\underline m,D):= {1\over \Gamma(\omega)(2i\pi)^n\pi^{lD\over2}} \, \int_{\IR^{(1,D-1)L}}
  e^{-i\sum_{i=1}^n x_i(\ell_i^2+m_i^2-i\varepsilon)}
  \prod_{i=1}^ld^D\ell_i \prod_{i=1}^n  dx_i\,.
\end{equation}

At this stage the integral is similar to the one leading to the
parametric representation with the replacement $x_r\to i x_r$ with
$x_r\in \mathbb R$. Setting $\tilde x_r=ix_r$ and performing
the Gaussian integrals over the loop momenta, we get
\begin{equation}
  \pi_n(\underline s,\underline \xi^2,D):= {1\over (2i\pi)^n} \, \int_{i\IR^n} {\widetilde\cU^{\omega-{D\over2}}\over
    \widetilde\cF^{\omega}}\prod_{i=1}^n \delta(1-\sum_{i=1}^n \tilde x_i) d\tilde x_i\,.
\end{equation}
using the projective nature of the integrand we have ${\widetilde\cU^{\omega-D/2}\over
    \widetilde\cF^{\omega}}=i^{-n} {\cU^{\omega-D/2}\over
    \cF^{\omega}}$ and the integral can be rewritten as the torus
integral
\begin{equation}\label{e:pidef}
  \pi_\Gamma(\underline s,\underline \xi^2,D):=  {1\over (2i\pi)^n} \, \int_{|x_1]=\cdots=|x_{n-1}|=1} {\cU^{\omega-D/2}\over \cF^{\omega}}\prod_{i=1}^{n-1}  dx_i\,.
\end{equation}
This integral shares the same integrand with the Feynman integral
$I_\Gamma$ in~\eqref{e:GraphPara2} but the cycle of integration
differs since we are integrating over a $n$-torus.
We show in \S\ref{sec:gzk-method} that this maximal cut arises naturally
from the toric formalism.

%-------------------------------------------------------------------------
\subsection{{\bf The differential equations }}

In general a Feynman integral $I_\Gamma(\underline s,\underline
\xi^2,\underline \nu,D)$ satisfies an inhomogeneous system of
differential equations
\begin{equation}\label{e:PFdef}
   \mathcal L_\Gamma I_\Gamma=\mathcal S_\Gamma,
\end{equation}
 where the inhomogeneous term $\mathcal S_\Gamma$
essentially arises from boundary terms corresponding to reduced
graph topologies where internal edges have been contracted.
Knowing the maximal cut integral allows to determine differential
operators $\mathcal L_\Gamma$
\begin{equation}
\mathcal  L_\Gamma \pi_\Gamma(\underline s,\underline \xi^2,D)=0  ,
\end{equation}
This fact has been exploited
in~\cite{Primo:2016ebd,Primo:2017ipr,Bosma:2017ens,Frellesvig:2017aai}  to obtain
the minimal order differential operator.
The important remark in this construction is to use that the only
difference between the Feynman integral $I_\Gamma$ and the maximal cut
$\pi_\Gamma$ is the choice of cycle of integration.
Since the Picard-Fuchs operator $\mathcal L_\Gamma$ acts as
\begin{equation}
\mathcal   L_\Gamma \pi_\Gamma(\underline s,\underline\xi^2,D)=
  \int_{\gamma_n} \mathcal L_\Gamma \Omega_F=
  \int_{\gamma_n}  d( \beta_\Gamma)=0
\end{equation}
this integral vanishes because the cycle $\gamma_n=\{|x_1|=\cdots=|x_n|=1\}$
has no boundaries $\partial \gamma_n=\emptyset$.
In the case of the Feynman integral $I_\Gamma$ this is not longer true
as
\begin{equation}
\mathcal  L_\Gamma I_\Gamma(\underline s,\underline\xi^2,D)=
    \int_{\Delta_n}  d(
    \beta_\Gamma)=\int_{\partial\Delta_n}
    \beta_\Gamma=\mathcal S_\Gamma\neq0\,.
\end{equation}
The boundary contributions arises from the configuration with some of
the Schwinger coordinate  $x_i=0$ vanishing which corresponds to the
so-called reduced topologies that are known to arises when applying
the integration-by-part algorithm (see~\cite{Chetyrkin:1981qh,Tarasov:1997kx,Tarasov:2017yyd} for instance).

We illustrate this logic on some elementary examples of differential equations for multi-valued
integrals relevant for the one- and two-loop massive sunset integrals
discussed in this text.

\subsubsection{The logarithmic integral}
\label{sec:bubble1}

We consider the integral
\begin{equation}
I_1(t)=\int_{a}^b {dx\over x(x-t)}  \,,
\end{equation}
and its cut integral
\begin{equation}
  \pi(t)=   \int_{\gamma} {dx\over x(x-t)}  \,,
\end{equation}
where $\gamma$ is a cycle around the point $x=t$.
Clearly we have
\begin{equation}
  {d\over dx}\left( 1\over t-x\right)= {1\over x(x-t)}+t{d\over
    dt}\left( 1\over x(x-t)\right) \,,
\end{equation}
therefore the integral $\pi(t)$ satisfies the differential equation
\begin{equation}
  t{d\over dt} \pi(t)+\pi(t)=  \int_{\gamma}     {d\over dx}\left( 1\over t-x\right)=0\,,
\end{equation}
and the integral $I_1(t)$ satisfies
\begin{equation}
  t{d\over dt} I_1(t)+I_1(t)= \int_{a}^b     {d\over dx}\left( 1\over
    t-x\right)={1\over b(b-t)}-{1\over a(a-t)}\,.
\end{equation}
Changing variables from $t$ to $p^2$ or an internal mass will  give the familiar differential
equation for the one-loop bubble that will be commented further in \S\ref{sec:bubble-graph}.

\subsubsection{ Elliptic curve}
\label{sec:ellipticcurve}

The second example is the differential equation for the period of an
elliptic curve $\mathcal E : y^2z=x(x-z)(x-tz)$ which is the geometry
of the two-loop sunset integral.  Consider the
differential of the first kind on the elliptic curve
\begin{equation}
    \omega= {dx\over \sqrt{x(x-1)(x-t)}}\,,
\end{equation}
this form can be seen as a residue evaluated on the elliptic curve
$\omega=\textrm{Res}_{\mathcal E}  \Omega$ of the form on the
projective space $\mathbb P^2$
\begin{equation}
    \Omega=   {  \Omega_0\over y^2z-x(x-z)(x-tz)}\,.
\end{equation}
where $\Omega_0= z dx\wedge dy+y dz\wedge dx+x
     dy\wedge dz$ is the natural top form on the projective space $[x:y:z]$.
Systematic ways of deriving  Picard-Fuchs operators for elliptic
curve is given by Griffith's algorithm~\cite{Griffiths1960}. Consider the
second derivative with respect to the parameter $t$

\begin{equation}
  {d^2\over dt^2} \Omega=   2{x^2(x-z)^2z^2\over (y^2z-x(x-z)(x-tz))^2}\Omega_0
\end{equation}
the numerator belongs to the Jacobian ideal\footnote{An ideal $I$ of a
  ring $R$, is the subset $I\subset R$, such that 1) $0\in I$, 2) for
  all $a, b\in I$ then $a+b\in I$, 3) for $a\in I$ and $b\in R$,
  $a\cdot b\in R$. For $P(x_1,\dots,x_n)$ an homogeneous polynomial in
  $R=\mathbb C[x_1,\dots,x_n]$ the Jacobian ideal of $P$ is the ideal
  generated by the first partial derivative  $\{\partial_{x_i}
  P(x_1,\dots,x_n)\}$~\cite{CoxKatz}. Given a multivariate polynomial
  $P(x_1,\dots,x_n)$ its Jacobian ideal is easily evaluated using {\tt
    Singular} command {\tt jacob(P)}. The hypersuface
  $P(x_1,\cdots,x_n)=0$ for an homogeneous polynomial, like the
  Symanzik polynomials, is of codimension 1 in the projective space
  $\mathbb P^{n-1}$. The singularities of the hypersurface are
  determined by the irreducible factors of the polynomial. This
 determines the  cohomology of the complement of the graph
 hypersurface and the number of independent master integrals
  as shown in~\cite{Bitoun:2017nre}.} of the polynomial $p(x,y,z):=y^2z-x(x-z)(x-tz)$,
$J_1=\langle \partial_x p(x,y,z)=-3x^2+2(t+1) xz-t z^22, \partial_y
p(x,y,z)=2yz,\break\partial_z p(x,y,z)=(t+1)x^2+y^2-2txz \rangle$,
since
\begin{equation}
  x^2(x-z)^2z^2= m^1_x \partial_x p(x,y,z)+ m^1_y \partial_y
  p(x,y,z)+m^1_z \partial_z p(x,y,z)\,.
\end{equation}
This implies that
\begin{multline}
     {d^2\over dt^2} \Omega ={\partial_x m^1_x+\partial_y m^1_y+\partial_z
       m^1_z\over (y^2z-x(x-z)(x-tz))^2}\Omega_0\cr
+ d\left((ym^1_z-z m^1_y)dx+(zm^1_x-xm^1_z)dy+(xm^1_y-ym^1_x)dz\over  (y^2z-x(x-z)(x-tz))^2\right)
\end{multline}
therefore
\begin{multline}
     {d^2\over dt^2} \Omega  + p_1(t) {d\over dt}\Omega={-p_1(t)x(x-z)z+\partial_x m^1_x+\partial_y m^1_y+\partial_z
       m^1_z\over (y^2z-x(x-z)(x-tz))^2}\Omega_0\cr
+ d\left((ym^1_z-z m^1_y)dx+(zm^1_x-xm^1_z)dy+(xm^1_y-ym^1_x)dz\over
  (y^2z-x(x-z)(x-tz))^2\right)\,.
\end{multline}
One easily derives that $\partial_x m^1_x+\partial_y m^1_y+\partial_z
       m^1_z$ is in the Jacobian ideal generated by $J_1$ and $x(x-z)z$
with the result that
\begin{multline}
 \partial_x m^1_x+\partial_y m^1_y+\partial_z
       m^1_z=    m^2_x \partial_x p(x,y,z)+ m^2_y \partial_y
  p(x,y,z)+m^2_z \partial_z p(x,y,z)\cr+ {2t-1\over t(t-1)} x(x-z)z,
\end{multline}
therefore  $p_1(t)= {2t-1\over t(t-1)}$ and  the Picard-Fuchs operator reads
\begin{multline}
     {d^2\over dt^2} \Omega +  {2t-1\over t(t-1)} {d\over dt}\Omega-{\partial_x m^2_x+\partial_y m^2_y+\partial_z
       m^2_z\over (y^2z-x(x-z)(x-tz))^2}\Omega_0=\cr
 d\left((ym^1_z-z m^1_y)dx+(zm^1_x-xm^1_z)dy+(xm^1_y-ym^1_x)dz\over
  (y^2z-x(x-z)(x-tz))^2\right)\cr
+d\left((ym^2_z-z m^2_y)dx+(zm^2_x-xm^2_z)dy+(xm^2_y-ym^2_x)dz\over
  y^2z-x(x-z)(x-tz)\right)\,.
\end{multline}
since  $\partial_x m^2_x+\partial_y m^2_y+\partial_z
       m^2_z=-{1\over 4t(t-1)}$ we have that

\begin{equation}
\left(  4t(t-1) {d^2\over dt^2}- 4(2t-1) {d\over dt} +1\right)\omega =
  -2 \partial_x \left( y\over  (x-t)^2\right)\,.
\end{equation}
For $\alpha$ and $\beta$ a (sympletic) basis of $H_1(\mathcal
E,\mathbb Z)$ the period integrals $\varpi_1(t):=\int_{\alpha} \omega$ and
$\varpi_2(t):=\int_{\beta}\omega$ both satisfy the differential
equation
\begin{equation}
\left(  4t(t-1) {d^2\over dt^2}- 4(1-2t) {d\over dt} +1\right)\varpi_i(t) =0\,.
\end{equation}

Again this differential operator acting on an integral with a
different domain of integration can lead to an homogeneous terms as
this is case for the two-loop sunset Feynman integral.

\medskip

The all procedure is easily implemented in {\tt Singular}~\cite{DGPS}
with the following set of commands

    \begin{Verbatim}[commandchars=\\\{\}]
{\color{incolor}In [{\color{incolor}1}]:} // Griffith-Dwork method for
deriving the Picard-Fuchs operator for the elliptic curve
 y\PYZca{}2z=x(x\PYZhy{}z)(x\PYZhy{}tz)
\end{Verbatim}

    \begin{Verbatim}[commandchars=\\\{\}]
{\color{incolor}In [{\color{incolor}2}]:} ring A=(0,t),(x,y,z),dp;
\end{Verbatim}

    \begin{Verbatim}[commandchars=\\\{\}]
{\color{incolor}In [{\color{incolor}3}]:} poly f=y\PYZca{}2*z\PYZhy{}x*(x\PYZhy{}z)*(x\PYZhy{}t*z);
\end{Verbatim}

    \begin{Verbatim}[commandchars=\\\{\}]
{\color{incolor}In [{\color{incolor}4}]:} ideal I1=jacob(f); I1
\end{Verbatim}

\begin{Verbatim}[commandchars=\\\{\}]
{\color{outcolor}Out[{\color{outcolor}4}]:} I1[1]=-3*x2+(2t+2)*xz+(-t)*z2
         I1[2]=2*yz
         I1[3]=(t+1)*x2+y2+(-2t)*xz
\end{Verbatim}

    \begin{Verbatim}[commandchars=\\\{\}]
{\color{incolor}In [{\color{incolor}5}]:} matrix M1=lift(I1,x\PYZca{}2*(x\PYZhy{}z)\PYZca{}2*z\PYZca{}2); M1
\end{Verbatim}

\begin{Verbatim}[commandchars=\\\{\}]
{\color{outcolor}Out[{\color{outcolor}5}]:} M1[1,1]=2/(3t+3)*xz3
         M1[2,1]=-1/(2t+2)*x2yz+1/(6t+6)*yz3
         M1[3,1]=1/(t+1)*x2z2-1/(3t+3)*z4
\end{Verbatim}

    \begin{Verbatim}[commandchars=\\\{\}]
{\color{incolor}In [{\color{incolor}6}]:} // checking the decomposition
x\PYZca{}2*(x\PYZhy{}z)\PYZca{}2*z\PYZca{}2\PYZhy{}M1[1,1]*I[1]\PYZhy{}M1[2,1]*I[2]\PYZhy{}M1[3,1]*I[3]
\end{Verbatim}

\begin{Verbatim}[commandchars=\\\{\}]
{\color{outcolor}Out[{\color{outcolor}6}]:} 0
\end{Verbatim}

    \begin{Verbatim}[commandchars=\\\{\}]
{\color{incolor}In [{\color{incolor}7}]:}  poly dC1=diff(M[1,1],x)+diff(M[2,1],y)
+diff(M[3,1],z);
dC1
\end{Verbatim}

\begin{Verbatim}[commandchars=\\\{\}]
{\color{outcolor}Out[{\color{outcolor}7}]:}  dC1=3/(t+1)*x2z-1/(t+1)*z3
\end{Verbatim}

    \begin{Verbatim}[commandchars=\\\{\}]
{\color{incolor}In [{\color{incolor}8}]:}  ideal I2=jacob(f),x*(x\PYZhy{}z)*z;
\end{Verbatim}

    \begin{Verbatim}[commandchars=\\\{\}]
{\color{incolor}In [{\color{incolor}9}]:}  matrix M2=lift(I2,dC1); M2
\end{Verbatim}

\begin{Verbatim}[commandchars=\\\{\}]
{\color{outcolor}Out[{\color{outcolor}9}]:} M2[1,1]=1/(2t2+2t)*z
         M2[2,1]=1/(4t2-4t)*y
         M2[3,1]=-1/(2t2-2t)*z
         M2[4,1]=(2t-1)/(t2-t)
\end{Verbatim}

    \begin{Verbatim}[commandchars=\\\{\}]
{\color{incolor}In [{\color{incolor}10}]:} // checking the decomposition
dC1\PYZhy{}M2[1,1]*I[1]\PYZhy{}M2[2,1]*I[2]\PYZhy{}M2[3,1]*I[3]
\PYZhy{}M2[4,1]*x*(x\PYZhy{}z)*z
\end{Verbatim}

\begin{Verbatim}[commandchars=\\\{\}]
{\color{outcolor}Out[{\color{outcolor}10}]:} 0
\end{Verbatim}

    \begin{Verbatim}[commandchars=\\\{\}]
{\color{incolor}In [{\color{incolor}11}]:}  poly dC2=diff(M2[1,1],x)+diff(M2[2,1],y)
+diff(M2[3,1],z);
dC2
\end{Verbatim}

\begin{Verbatim}[commandchars=\\\{\}]
{\color{outcolor}Out[{\color{outcolor}11}]:} -1/(4t2-4t)
   \end{Verbatim}

%%%%%%%%%%%%%%%%%%%%%%%%%%%%%%%%%%%%%%%%%%%%%%%%%%%%%%%%%%%%%%%%%%
\section{{\bf Toric geometry and Feynman graphs}}\label{sec:toric}

We will show how the toric approach provides a nice way to obtain this
maximal cut integral.
The maximal cut integral $\pi_\Gamma(\underline s,\underline \xi^2,D)$
is the particular case of generalised Euler integrals
\begin{equation}
  \int_\sigma \prod_{i=1}^r P_i(x_1,\dots,x_n)^{\alpha_i}
  \prod_{i=1}^n x_i^{\beta_i} dx_i
\end{equation}
studied by Gel'fand, Kapranov and Zelevinski (GKZ)
in~\cite{GKZ1,GKZ2}. There $P_i(x_1,\dots,x_n)$ are  Laurent
polynomials,  $\alpha_i$ and $\beta_i$ are complex numbers and
$\sigma$ is a cycle.  The cycle entering the maximal cut integral
in~\eqref{e:pidef} is the product of circles
$\sigma=\{|x_1|=|x_2|=\cdots=|x_n|=1\}$. But other cycles arise when
considering different cuts of Feynman graphs.
The GKZ approach provides a totally combinatorial approach to
differential equation satisfied by these integrals.

As well in the case when $P(\underline x,\underline z)=\sum_{i}
z_{i_1,\dots,i_r} \prod_{i=1}^n x_i^{\alpha_i}$ is the Laurent
polynomial defining a Calabi-Yau hypersurface $\{P(\underline
x,\underline z)=0\}$,
Batyrev showed that there is one canonical period integral~\cite{Batyrev:1993jq,COX:1994bt}
\begin{equation}
  \Pi(\underline z):={1\over(2i\pi)^n} \int_{|x_1|=\cdots=|x_n|=1}
  {1\over P(\underline
x,\underline z)}  \prod_{i=1}^n {dx_i\over x_i}\,.
\end{equation}
This corresponds to the maximal cut integral~\eqref{e:pidef} In the
case where $\omega=D/2=1$ which is satisfied by the $(n-1)$-loop
sunset integral $D=2$ dimensions.  The graph hypersurface of the
$(n-1)$-loop sunset (see~\eqref{e:Fsunsetdef})  is always a
Calabi-Yau $(n-1)$-fold. See for more comments about this at the end
of \S\ref{sec:suns-graph-polyn}.  We refer to the
reviews~\cite{Hosono:1994av,Closset:2009sv}  for some introduction to
toric geometry for physicts.

\subsection{Toric polynomials and Feynman graphs}
\label{sec:toric-polynomials-feyn}

The second Symanzik polynomial $\mathcal F(\underline s,\underline
\xi^2,x_1,\dots,x_n)$ defined in~\eqref{e:Fdef} is a specialisation of
the homogeneous (toric) polynomial\footnote{Consider an homogeneous polynomial of degree $d$ $$P(\underline z,\underline x)=\sum_{0\leq r_i\leq n\atop r_1+\cdots +r_n=d}z_{i_1,\dots,i_n}\prod_{i=1}^n x_i^{r_i}$$ this is called a \emph{toric polynomial} if it is invariant under the following actions
$$
z_i \to \prod_{j=1}^n t_i^{\alpha_{ij}} z_i; \qquad x_i\to \prod_{j=1}^n t_i^{\beta_{ij}} x_i
$$
for $(t_1,\dots,t_n)\in \mathbb C^n$ and  $\alpha_{ij}$ and
$\beta_{ij}$  integers. The second Symanzik polynomial have a natural
torus action acting on the mass parameters and the kinematic variables
as we will see on some examples below. We refer to the book~\cite{CoxKatz} for more details.  } of degree $l+1$ at most quadratic in each
variables in  the
projective variables $(x_1,\dots,x_n) \in\mathbb P^{n-1}$
\begin{equation}\label{e:Ftoric}
\mathcal F^{toric}_l(\underline z,x_1,\dots,x_n)=\mathcal
U^{tor}_l(x_1,\dots,x_n)\left(\sum_{i=1}^n \xi_i^2 x_i\right)-\mathcal V^{tor}_l(x_1,\dots,x_n),
\end{equation}
where  for $l\leq n$
\begin{equation}
  \label{e:Utor}
  \mathcal
U^{tor}_l(x_1,\dots,x_n) :=\sum_{0\leq r_i\leq
  1\atop r_1+\cdots +r_n=l}
u_{i_1,\dots,i_n}\prod_{i=1}^n x_i^{r_i}   ,
\end{equation}
where the coefficients  $u_{i_1,\dots,i_n}\in \{0,1\}$.
The expression in~\eqref{e:Ftoric} is the most generic form compatible
with the properties of the Symanzik polynomials listed in \S\ref{sec:feynm-param}.

\medskip

There are
${n!\over (n-l)!l!}$ independent coefficient in the polynomial $\mathcal
U^{tor}_l(x_1,\dots,x_n) $. Of course
this is a huge over counting, as this does not take into account the
symmetries of the graphs and the constraints on the non-vanishing of
some coefficients. This will be enough for the toric description we
are using here.  In order to keep most of the combinatorial  power of the
toric approach we will only do the specialisation of the toric
coefficients with the physical slice corresponding of Feynman graph
polynomial at the end on solutions. This will avoid having to think at constrained
system of differential equations which is a difficult problem
discussed recently in~\cite{Bitoun:2017nre}.

The kinematics part has the toric polynomial
\begin{equation}\label{e:Vtor}
  \mathcal V^{tor}_l(x_1,\dots,x_n):=\sum_{0\leq r_i\leq
  1\atop r_1+\cdots +r_n=l+1}
z_{i_1,\dots,i_n}\prod_{i=1}^n x_i^{r_i}    ,
\end{equation}
where the coefficients $z_{i_1,\cdots,i_n}\in \mathbb C$.
The number of independent toric variables $z_{\underline i}$ in $\mathcal
V^{tor}(x_1,\dots,x_n)$  is ${n!\over (n-l-1)!(l+1)!}$.

\subsubsection{Some important special cases}
\label{sec:some-special-cases}

There are few important special cases.
\begin{itemize}
\item
At one-loop order $l=1$ and the number of  independent toric
variables in $\mathcal V^{tor}(x_1,\dots,x_n)$ is exactly the number
of independent kinematics for an $n$-gon one-loop amplitude
$$\includegraphics[width=4cm]{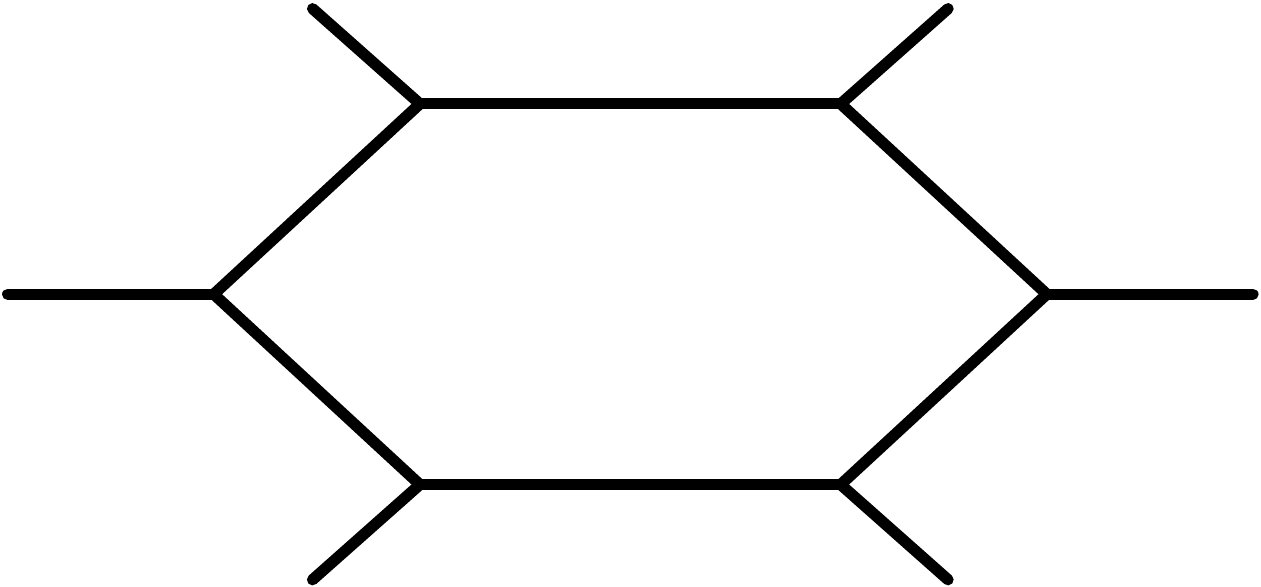}\,.$$
In this case the most general toric one-loop polynomial is

\begin{equation}\label{e:tor1}
    \mathcal F^{tor}_1(x_1,\cdots,x_n)= \left(\sum_{i=1}^n x_i\right)
    \left(\sum_{i=1}^n \xi_i^2 x_i\right)-\mathcal V^{tor}_1(x_1,\cdots,x_n).
\end{equation}

\item For $l=n$ there is only one vertex the graph is  $n$-bouquet which is
a product of $n$ one-loop graphs. These graphs contribute to the
reduced topologies entering the determination of the inhomogeneous
term $\mathcal S_\Gamma$ of the Picard-Fuchs equation~\eqref{e:PFdef}. They
don't contribute to the maximal cut $\pi_\Gamma$ for $l>1$.

$$
\includegraphics[width=4cm]{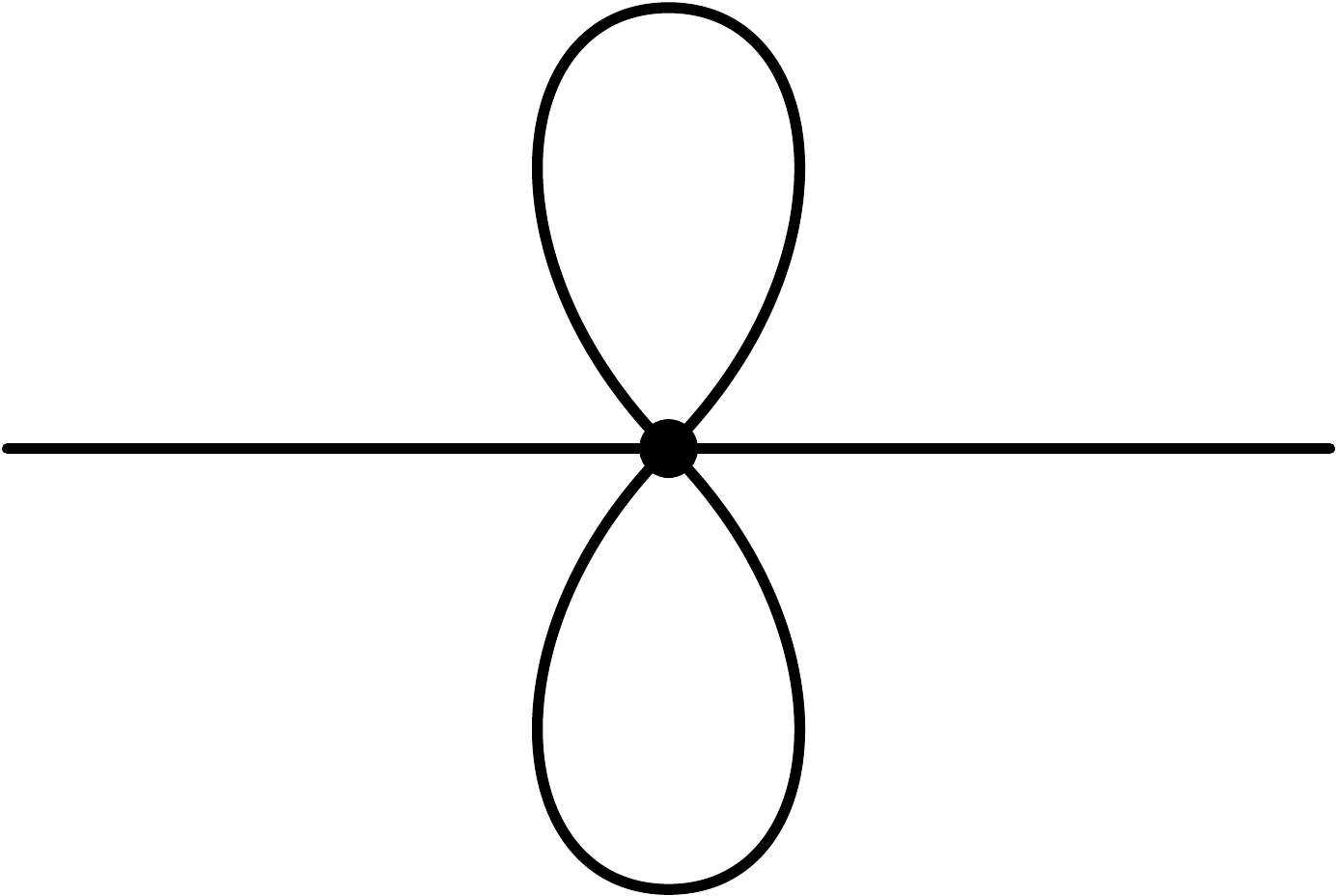}
$$

\item The case $l=n-1$ corresponds to the $(n-1)$-loop  two-point
  sunset graphs
$$\includegraphics[width=4cm]{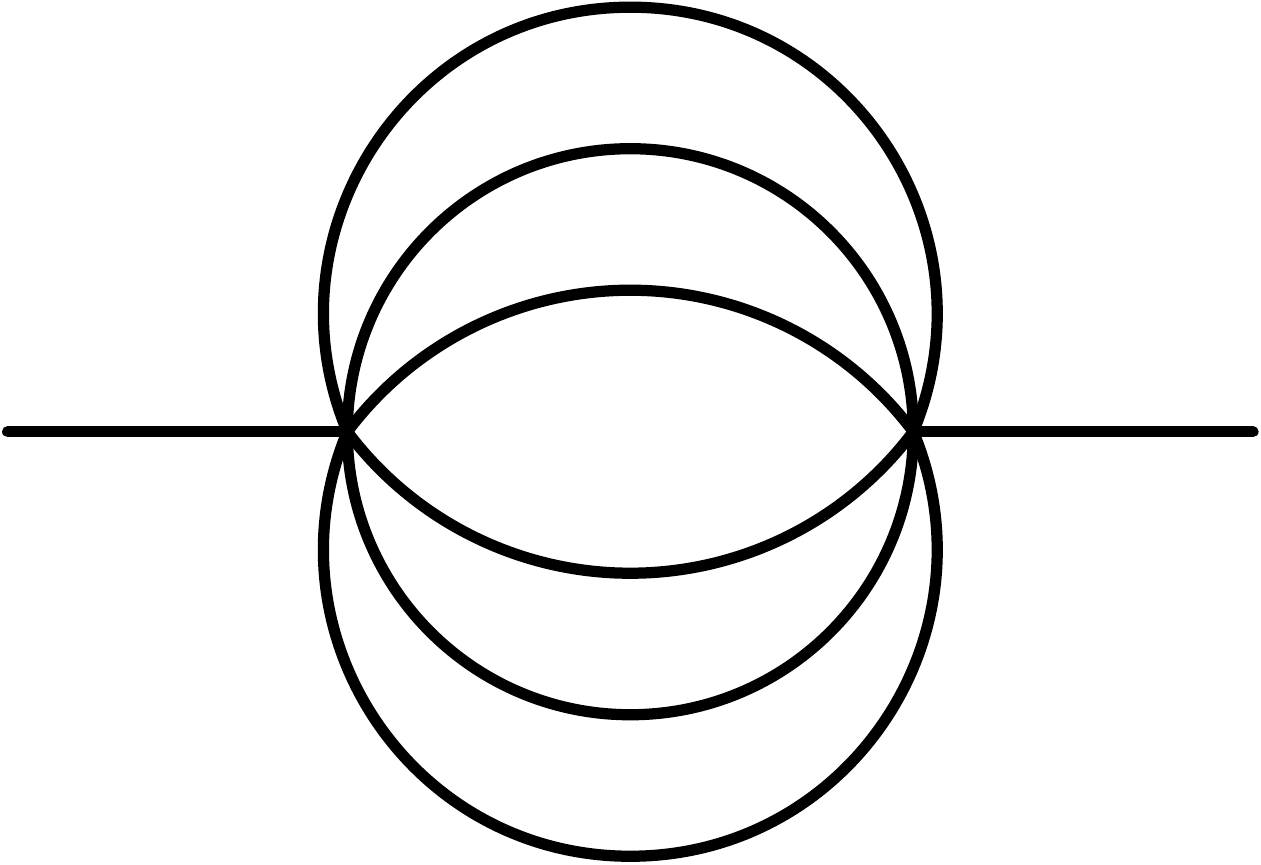}\,.$$
  In that case the kinematic polynomial is just
\begin{equation}
  \mathcal V^{tor}_{n-1}(x_1,\dots,x_n)= z_{1,\dots,1} x_1\cdots x_n  ,
\end{equation}
and the toric polynomial
\begin{equation}
\mathcal F^{tor}_{n-1}(x_1,\dots,x_n)=x_1\cdots x_n\left(\sum_{i=1}^n
  {u_{1,\dots,0,\dots,1}\over x_i}\right)  \left(\sum_{i=1}^{n}\xi_i^2x_i\right)-z_{1,\dots,1} x_1\cdots x_n  ,
\end{equation}
where the index $0$ in $u_{1,\dots,0,\dots,1}$ is at position $i$.
Actually by redefining the parameter $z_{1,\dots,1}$ the generic toric
polynomial associated to the sunset graph are
\begin{equation}
\mathcal F^{tor}_\su(x_1,\dots,x_n)=x_1\cdots x_n\left(\sum_{1\leq
    i,j\leq n\atop i\neq j} z_{ij} {x_i\over x_j} -z_{0}\right),
\end{equation}
where $z_{ij}\in\mathbb C$ and $z_0\in\mathbb C\backslash\{0\}$. This
polynomial   has $1-n+n^2$ parameters where a sunset graph as $n+1$
physical  parameters given by $n$ masses and one kinematics invariant
\begin{equation}
  \label{e:Fsunsetdef}
\mathcal F_\su^l(p^2,\underline\xi^2,\underline x)= x_1\cdots x_{l+1}\left(\sum_{i=1}^{l+1}
  {1\over x_i}\right)  \left(\sum_{i=1}^{l+1}\xi_i^2x_i\right)-p^2 x_1\cdots x_{l+1}\,.
\end{equation}
 So there are too many parameters from $n\geq3$ but this
generalisation will be useful for the GKZ description used in the next sections.
\end{itemize}

%------------------------------------------------------------------------
\subsection{The GKZ approach : a review}
\label{sec:gzk-method}

In the section we briefly review the GKZ construction based
on~\cite{GKZ1,GKZ2} see as well~\cite{Stienstra:tf}.
We consider the Laurent  polynomial of $n-1$ variables $P(z_1,\dots,z_r)=\mathcal
F^{tor}\underline z,x_1,\dots,x_{n}/(x_1\cdots x_n)$ from the toric
polynomial of \S\ref{sec:toric-polynomials-feyn}.  The coefficients of monomials are $z_i$ (by
homogeneity  we set $x_n=1$)
\begin{equation}\label{e:PtoricDef}
P (z_1,\dots,z_r)=\sum_{\mathbf a=(a_1,\dots,a_{n-1})\in \mathbf A} z_{\pmb a}
\prod_{i=1}^{n-1} x_i^{a_i},
\end{equation}
with $\mathbf a=(a_1,\dots,a_{n-1})$ is an element of  $\mathbf A=(\mathbf
a_1,\dots,\mathbf a_r)$ a finite subset of
$\mathbb Z^{n-1}$. The number of elements in $A$ is $r$ the
number of monomials in $P (z_1,\dots,z_r)$.

We consider the natural fundamental period integral~\cite{Batyrev:1993wa}
\begin{equation}\label{e:Pidef}
  \Pi(\underline z):={1\over(2i\pi)^{n-1}}\int_{|x_1|=\cdots=|x_{n-1}|=1}   P(z_1,\dots,z_r)^m
\prod_{i=1}^{n-1} {dx_i\over x_i}  ,
\end{equation}
 which is the same as maximal cut
$\pi_\Gamma$ in~\eqref{e:pidef} for  $D=2\omega=-m$.
The derivative with respect to $z_{\mathbf a}$ reads
\begin{equation}
  {\partial\over\partial z_{\pmb a}}\Pi(\underline z)=
  {1\over(2i\pi)^{n-1}}\int_{|x_1|=\cdots=|x_{n-1}|=1} \,  m   P(z_1,\dots,z_r)^{m-1}
\prod_{i=1}^{n-1} x_i^{a_i}\,{dx_i\over x_i}  ,
\end{equation}
therefore for every vector $\pmb \ell=(\ell_1,\dots,\ell_r)\in \mathbb Z^{n-1}$ such that
\begin{equation}
\ell_1+\cdots+\ell_r=0, \qquad \ell_1\mathbf a_1+\cdots +\ell_r
\mathbf a_r= \pmb \ell\cdot \mathbf A =0,
\end{equation}
then  holds the differential equation

\begin{equation}
\left(\prod_{l_i>0} \partial_{z_i}^{l_i}
-\prod_{l_i<0} \partial_{z_i}^{-l_i} \right)   \,\Pi(\underline z)
=0.
\end{equation}

Introducing the
so-called $\mathcal A$-hypergeometric functions\footnote{The convergence
of these series is   discussed in \cite[\S3-2]{Hosono:1998ct}
and~\cite[\S5.2]{Stienstra:tf}.} $\Phi_{\mathbb
  L,\gamma}(z_1,\dots,z_r)$  of $r$ complex variables
$(z_1,\dots,z_r)\in\mathbb C^r$
\begin{equation}
\Phi_{\mathbb L,\pmb\gamma}(z_1,\dots,z_r)=
\sum_{(\ell_1,\dots,\ell_r)\in\mathbb L}\prod_{j=1}^r
{z_j^{\gamma_j+\ell_j}\over \Gamma(\gamma_j+\ell_j+1)},
\end{equation}
depending on the complex parameters
$\pmb\gamma:=(\gamma_1,\dots,\gamma_r)\in\mathbb C^r$ and
the lattice
\begin{equation}
   \mathbb L:=\{(\ell_1,\dots,\ell_r)\in\mathbb Z| \sum_{i=1}^r \ell_i
\mathbf a_i=0,  \ell_1+\cdots+\ell_r=0\},
 \end{equation}
with  $r$ elements $\{\mathbf
a_1,\dots,\mathbf a_r\}\in
\mathbb Z^n$.   These functions are  solutions of the so-called $\mathcal A$-hypergeometric system of
differential equations given by
 a vector $\mathbf c\in \mathbb C^n $ and :
\begin{itemize}
\item For every $\pmb \ell=(\ell_1,\dots,\ell_r)\in
\mathbb L$  there is one
  differential operator
\begin{equation}
\Box_{\pmb \ell}:=\prod_{\ell_i>0} \partial_{z_i}^{\ell_i}
-\prod_{\ell_i<0} \partial_{z_i}^{-\ell_i} ,
\end{equation}
such that $\Box_{\pmb \ell} \Phi_{\mathbb
  L,\gamma}(z_1,\dots,z_r)=0$

\item $n$ differential operators $\mathbf E:=(E_1,\dots,E_{n-1})$
\begin{equation}
\mathbf E:= \mathbf a_1 z_1{\partial\over\partial z_1}+\cdots+\mathbf a_r
z_r{\partial\over\partial z_r},
\end{equation}
such that for $\mathbf c=(c_1,\dots,c_{n-1})$ we have
\begin{equation}
(\mathbf E-\mathbf c) \Phi_{\mathbb
  L,\gamma}(z_1,\dots,z_r)=0.
\end{equation}
Notice that $E_1= \sum_{i=1}^n z_i{\partial\over\partial z_i}$ is the
Euler operator and $c_1$ is the degree of homogeneity of the
hypergeometric function.

These operators satisfy the commutation relations
\begin{align}
\mathbf z^{\mathbf
    u}  \mathbf E -\mathbf E \mathbf z^{\mathbf
    u} &= - (\mathbf A\cdot \mathbf u)\,  \mathbf z^{\mathbf
    u},\cr
\partial_z^{\mathbf
    u}  \mathbf E -\mathbf E \partial_z^{\mathbf
    u} &=  (\mathbf A\cdot \mathbf u)\,  \partial_z^{\mathbf
    u},
\end{align}
with $\mathbf z^{\mathbf
    u}:=\prod_{i=1}^r z_r^{u_r}$ and $\partial_z^{\mathbf
    u} :=\prod_{i=1}^r \partial_{z_r}^{u_r}$.
\end{itemize}

Using the GKZ construction one can easily derive a system of
differential operator annihilating the maximal of any Feynman
integral after identification of the toric variables with the physical
parameters.  The system of differential operators obtained from the
GKZ system can be massaged into a set of Picard-Fuchs differential
operators in a spirit similar to the one used in mirror
symmetry~\cite{Hosono:1993qy,Hosono:1994ax,CoxKatz}.

Since it is rather complicated to restrict differential operators but
it is easier to restrict functions, it is therefore preferable to
determine the $\mathcal A$-hypergeometric representation of the maximal cut integral and derive the minimal
differential operator annihilating this integral.
For well chosen vector $\pmb \ell\in\mathbb L$ the
differential operator factorises with a factor being given by the
minimal (Picard-Fuchs) differential operator acting on the Feynman
integral.

An important remark is that the  maximal cut integral
\begin{equation}
 \pi_{\Gamma}= \int_{|x_1|=\cdots=|x_{n-1}|=1}  {1\over \mathcal F_\Gamma}
\prod_{i=1}^{n-1} dx_i,
\end{equation}
is a particular case of fundamental period $\Pi(\underline z)$ in~\eqref{e:Pidef} with $m=-1$ and therefore is given by a $\mathcal
A$-hypergeometric function once we have identified the toric variables
$z_i$ with the physical parameters.

In the next section we illustrate this approach on some simple but
fundamental examples.

\subsection{Hypergeometric functions and GKZ system}
\label{sec:hyperg-funct-gkz}

The relation between hypergeometric functions and the GKZ differential
system can be simply understood as follows
(see~\cite{Cattani,Beukers:2016er,Stienstra:tf}).

\subsubsection{The Gau\ss{} hypergeometric series}
\label{sec:gauss-hyperg-seri}

Consider the case of $\mathbb L=(1,1,-1,1)\mathbb Z \subset \mathbb
Z^4$ and the vector $\gamma=(0,c-1,-a,-b)\in\mathbb C^4$ and $c$ a
positive integer. The GKZ hypergeometric function is
\begin{equation}
  \Phi_{\mathbb L,\gamma}(u_1,\dots,u_4)= \sum_{n\in \mathbb Z} {u_1^n
    u_2^{1-c+n} u_3^{-a-n} u_4^{-b-n}\over \Gamma(1+n)
    \Gamma(c+n)\Gamma(1-n-a)\Gamma(1-n-b)}\,,
\end{equation}
which can be rewritten as
\begin{equation}
 \Phi_{\mathbb L,\gamma}(u_1,\dots,u_4)= {u_2^{c-1}u_3^{-a}u_4^{-b}\over \Gamma(c)\Gamma(1-a)\Gamma(1-b)}
  \,
_2F_1\left({a,\ b\atop c}\Big| {u_1u_2\over u_3u_4}\right)\,.
\end{equation}
The GKZ system is
\begin{align}
\left(  {\partial^2\over \partial u_1\partial u_2}-
  {\partial^2\over \partial u_3\partial u_4}  \right)\Phi_{\mathbb
  L,\gamma}(u_1,\dots,u_4)&=0,\cr
\left(  u_1 {\partial\over \partial u_1}-u_2
  {\partial\over \partial u_2}  +1-c\right)\Phi_{\mathbb
  L,\gamma}(u_1,\dots,u_4)&=0,\cr
\left(  u_1 {\partial\over \partial u_1}+u_3
  {\partial\over \partial u_3}  +a\right)\Phi_{\mathbb
  L,\gamma}(u_1,\dots,u_4)&=0,\cr
\left(u_1  {\partial\over \partial u_1}+u_4 {\partial\over \partial u_4}  +b\right)\Phi_{\mathbb L,\gamma}(u_1,\dots,u_4)&=0\,.
\end{align}
By differentiating we find
\begin{align}
\left( u_2  {\partial^2\over\partial u_1\partial u_2}  -
  u_1{\partial^2\over\partial u_1^2} +c {\partial\over \partial
  u_1}\right) \Phi_{\mathbb L,\gamma}(u_1,\dots,u_4)&=0,\cr
\left(u_3u_4 {\partial^2\over\partial u_3\partial
  u_4}-\left(u_1{\partial\over\partial u_1}+a\right)\left(u_1{\partial\over\partial u_1}+b\right) \right) \Phi_{\mathbb L,\gamma}(u_1,\dots,u_4)&=0\,.
\end{align}
combining these equations one finds
\begin{multline}
\left(  u_1^2{\partial \over \partial
  u_1}+(1+a+b)u_1{\partial\over\partial u_1}+ab\right) \Phi_{\mathbb
  L,\gamma}(u_1,\dots,u_4)\cr
={u_3u_4\over u_2}\, \left(u_1{\partial^2\over\partial
    u_1^2}+c{\partial\over\partial u_1}\right) \Phi_{\mathbb
  L,\gamma}(u_1,\dots,u_4).
\end{multline}
Setting $F(z)=\Gamma(c)\Gamma(1-a)\Gamma(1-b)\Phi_{\mathbb
  L,\gamma}(z,1,1,1)$ gives that $F(z)={}_2F_1({a\, b\atop c}|z)$ satisfies the Gau\ss{}
hypergeometric differential equation
\begin{equation}\label{e:hyper2F1}
  z(z-1){d^2 F(z)\over dz^2}+((a+b+1)z+c){d F(z)\over dz}+ab F(z)=0\,.
\end{equation}

\subsection{The massive one-loop graph}
\label{sec:bubble-graph}

In this section we show how to apply the GKZ formalism on the one-loop
bubble integral

$$
\includegraphics[width=4cm]{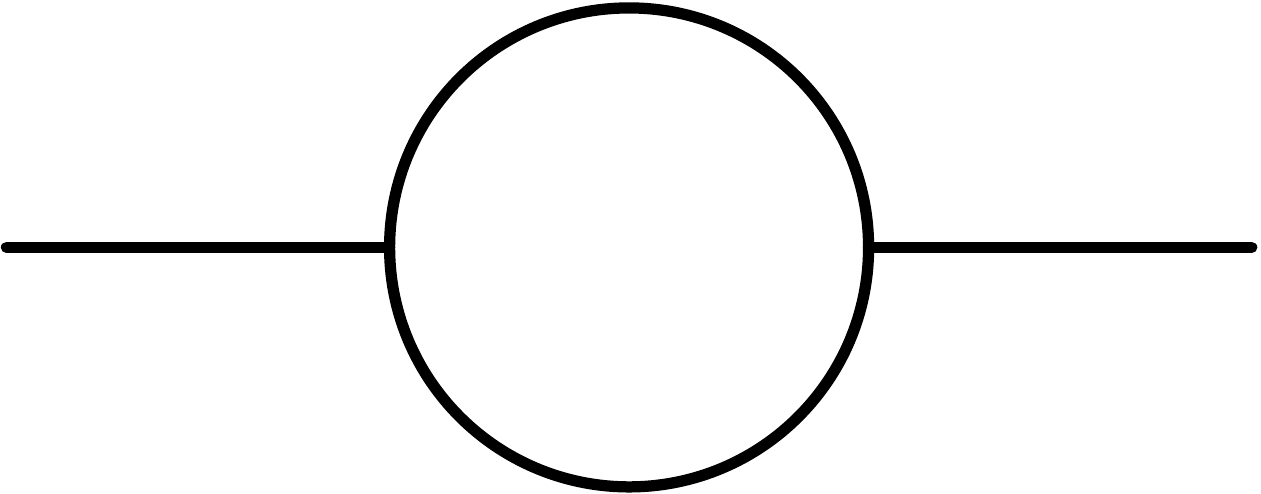}
$$

\subsubsection{Maximal cut}

The one-loop sunset (or bubble) graph as the graph polynomial
\begin{equation}\label{e:Fcirc}
  \mathcal F_\circ(x_1,x_2,t,\xi_1^2,\xi_2^2)=p^2x_1x_2 -(\xi_1^2x_1+\xi_2^2x_2)(x_1+x_2)\,.
\end{equation}
The most general toric degree two polynomial in $\mathbb P^2$ with at
most degree two monomial is given by
\begin{equation}\label{e:Ftor1}
\mathcal F^{tor}_\circ(x_1,x_2,z_1,z_2,z_3)=z_{1} x_1^2 + z_{2} x_2^2 + z_{3}x_1x_2\,.
\end{equation}
This toric polynomial has three parameters which is exactly the
number of independent physical parameters. The identification of the variables is
given by
\begin{equation}\label{e:physpar2}
z_1=-\xi_1^2, \qquad
z_2=-\xi_2^2, \qquad
z_3=p^2-(\xi_1^2  +\xi_2^2)\,,
\end{equation}
We consider the equivalent toric Laurent polynomial  
\begin{equation}
P(x_1,x_2) = {\mathcal F^{tor}_\circ\over x_1x_2}= \sum_{i=1}^3 z_i x_1^{a_i^1} x_2^{a_i^2} \,,
\end{equation}
so that $p^2$ in~\eqref{e:Fcirc} corresponds to the constant term (or
the origin the Newton polytope) 
and setting $\mathbf a_i= (1,a_i^1,a_i^2)$ we have
\begin{equation}
\mathbf  A_\circ= \begin{pmatrix}
    \mathbf a_1\cr \mathbf a_2\cr \mathbf a_3
  \end{pmatrix}=
  \begin{pmatrix}
 1 & -1 & 1 \cr
1 & 1 & -1 \cr
1 & 0 & 0
  \end{pmatrix}\,.
\end{equation}
The lattice is defined by
\begin{equation}
\mathbb L_\circ:=\{\pmb \ell:=(\ell_1,\ell_2,\ell_3)\in \mathbb
Z^3|\ell_1\mathbf a_1+\ell_2\mathbf a_2+\ell_3\mathbf a_3=\pmb \ell\cdot \mathbf
A_\circ=0 \}\,.
\end{equation}
This means that the elements of $\mathbb L_\circ$ are in the kernel of
$\mathbf A_\circ$.
This lattice in $\mathbb Z^3$ has rank one
\begin{equation}
\mathbb L_\circ=(1,1,-2)
\mathbb Z\,.
\end{equation}
 Notice that all the elements automatically satisfy the
condition $\ell_1+\ell_2+\ell_3=0$.

\medskip
Because the rank is one the  GKZ system of differential equations is
given by
\begin{align}\label{e:GKZbubble1}
 e_1&:= \frac{\partial^{2}}{\partial z_1\partial z_2} -
  \frac{\partial^{2}}{(\partial z_{3})^{2}},\cr
d_1&:=\sum_{r=1}^3  z_{r} \frac{\partial}{\partial z_{r}},\cr
d_2&:=z_1{\partial\over \partial z_1}-z_2{\partial\over \partial
     z_2},
\end{align}
By construction for $\alpha\in\mathbb C$
\begin{align}
    e_1(\mathcal F^{tor}_\circ)^\alpha&=0,\cr
d_1 (\mathcal F_\circ^{tor})^\alpha&=\alpha\, (\mathcal F_\circ^{tor})^\alpha,
\end{align}
 and
 \begin{equation}
d_2 (\mathcal F_\circ^{tor})^\alpha = \frac12\,\left( \partial_{x_1}(x_1
(  \mathcal F_\circ^{tor})^\alpha)- \partial_{x_2}(x_2
(  \mathcal F_\circ^{tor})^\alpha)\right)\,,
 \end{equation}
therefore the action of the derivative $d_2$ vanishes on the integral but not
the integrand
\begin{equation}
  d_2 \int_\gamma   (\mathcal F_\circ^{tor})^\alpha = 0 \qquad
  \textrm{for}\qquad \partial \gamma=\emptyset\,.
\end{equation}

\medskip

The GKZ hypergeometric series is defined as  for
$\gamma_i\not\in\mathbb Z$
\begin{equation}
  \Phi_{\mathbb L,\pmb \gamma}^\circ= \sum_{\pmb \ell\in\mathbb
    L_\circ} \prod_{i=1}^3 {z_i^{l_i+\gamma_i}\over  \Gamma(l_i+\gamma_i+1)},
\end{equation}
in this sum we have $\pmb \ell= n (1,1,-2)$ with
$n\in\mathbb Z$, and
the condition $\sum_{i=1}^3 \gamma_i \mathbf a_i=(0,0,-1)$
which can be solved using $\pmb \gamma=(\gamma_1,\gamma_2,\gamma_3)=
\gamma   (1,1,-2)+(0,0,-1)$, leading to
\begin{equation}
  \Phi_{\mathbb L,\pmb \gamma}^\circ= {1\over z_3} \sum_{n\in\mathbb
    Z} {u_1^{n} \over
    \Gamma(n+\gamma+1)^2  \Gamma(-2n+\gamma)},
 \end{equation}
where we have introduced  the new toric coordinate
\begin{equation}
  u_1:={z_1z_2\over z_3^2}= \frac{\xi_1^2 \xi_2^2 }{\left(p^2-(\xi_1^2 +\xi_2^2)\right)^2}\,.
\end{equation}
This is the natural coordinate dictated by the invariance of the period
integral under the transformation
$(x_1,x_2)\to (\lambda x_1,\lambda x_2)$ and $(z_1,z_2,z_3)\to
(z_1/\lambda,z_2/\lambda,z_3/\lambda)$.

\medskip

This GKZ hypergeometric function is a combination of $_3F_2$
hypergeometric functions
\begin{multline}
     \Phi_{\mathbb L,\pmb \gamma}^\circ={1\over
       z_3^{1-2\gamma}}\Bigg({u_1^{\gamma-1}\over
         \Gamma(\gamma)\Gamma(\gamma+2)}\,  {}_3F_2\left({1, \,1-\gamma,\,1-\gamma
           \atop 1+\frac{\gamma}2,\, \frac32+\frac{\gamma}2}\Big|{1\over
           4u_1}\right)\cr
+{u_1^\gamma\over \Gamma(\gamma+1)^2} \, {}_3F_2\left({1, \,\frac12-\frac{\gamma}2,\,\frac12-\frac{\gamma}2
           \atop 1+\gamma,\, 1+\gamma}\Big|4u_1\Bigg)
\right)\,.
\end{multline}

For $\gamma=0$ the series is trivially zero
as the system is resonant and needs to be
regularised~\cite{Stienstra:1997,Hosono:1998ct} . The regularisation
is to use the functional equation for the $\Gamma$-function
$\Gamma(z)\Gamma(1-z)=\pi/\sin(\pi z)$ to replace the pole term by
\begin{equation}
  \lim_{\epsilon\to0} {\Gamma(\epsilon)\over \Gamma(-2n+\epsilon)}=
  \Gamma(1+2n), \qquad n\in\mathbb Z\backslash\{0\},
\end{equation}
and write the associated regulated period as
\begin{equation}
  \pi_\circ =\lim_{\epsilon\to0} {1\over z_3}\, \sum_{n\in \mathbb N}
  {u_1^{n} \Gamma(\epsilon)\over
    \Gamma(n+1)^2\Gamma(-2n+\epsilon)}\,,
\end{equation}
which is easily shown to be
\begin{align}\label{e:Pi1sunset}
  \pi_\circ(z_1,z_2,z_3)
&={1\over z_3}{}_2F_1\left({\frac12 \ 1\atop
      1}\Big|4u_1\right)={1\over \sqrt{z_3^2-4z_1z_2}}\,,\cr
&= {1\over \sqrt{(p^2-(\xi_1+\xi_2)^2)(p^2-(\xi_1-\xi_2)^2)}}\,.
\end{align}
This expression of course matches the expression for the maximal cut~\eqref{e:pidef}
integral  $\pi_\circ(p^2,\xi_1^2,\xi_2^2,2)$ in two dimensions

\begin{equation}
  \pi_\circ(p^2,\xi_1^2,\xi_2^2,2)= {1\over (2i\pi)^2}\int_{|x_1|=|x_2|=1} {dx_1dx_2\over
    \mathcal F_\circ (x_1,x_2)}\,.
\end{equation}

%--------------------------------------------------------------------------
\subsubsection{The differential operator}
\label{sec:diff-oper}

From the expression of the maximal cut $\pi_\circ$ in~\eqref{e:Pi1sunset} as an
hypergeometric series, which satisfies a second order differential
equation~\eqref{e:hyper2F1}, we can extract a differential operator
with respect to $p^2$ or the masses $\xi_i^2$ annihilating the  maximal cut.  This differential equation is not the
minimal one as it can be factorised leaving  minimal order
differential operators are annihilating the maximal cut are
such that  $ L_{PF,(1)}^\circ\pi_\circ(p^2,\xi_1^2,\xi_2^2)=0$ and $
L_{PF,(2)}^\circ\pi_\circ(p^2,\xi_1^2,\xi_2^2)=0$ with
\begin{equation}\label{e:LPF1bubble}
  L_{PF,(1)}^\circ=  p^2 {d\over dp^2} + {p^2 (p^2-\xi_1^2-\xi_2^2)\over (p^2-(\xi_1+\xi_2)^2)(p^2-(\xi_1-\xi_2)^2)}\,,
\end{equation}
and
\begin{equation}\label{e:LPF2bubble}
  L_{PF,(2)}^\circ=  \xi_1^2 {d\over d\xi_1^2} - {\xi_1^2 (p^2-\xi_1^2+\xi_2^2)\over (p^2-(\xi_1+\xi_2)^2)(p^2-(\xi_1-\xi_2)^2)}\,,
\end{equation}
with of course a similar operator with the exchange of $\xi_1$ and
$\xi_2$.
These operators do not annihilate the integrand but lead to total
derivatives

\begin{equation}\label{e:PF1}
L_{PF,(1)}^{\circ}  \,{1\over \mathcal F_\circ(\underline x,p^2,\underline\xi^2)}=
  \partial_{x_1}
  \left(p^2(2 \xi_2^2 -(p^2-(\xi_1^2 +\xi_2^2 )) x_1 ) \over (p^2-(\xi_1+\xi_2)^2)(p^2-(\xi_1-\xi_2)^2)\mathcal F_2(x_1,1,p^2,\underline
    \xi^2)\right)\,,
\end{equation}
and
\begin{equation}\label{e:PF2}
 L_{PF,(2)}^\circ \, {1\over \mathcal F_\circ(\underline x,p^2,\underline\xi^2)} =
\partial_{x_1}\left(((p^2-\xi_2^2)^2-\xi_1^2(p^2+\xi_2^2))x_1-\xi_2^2(p^2+\xi_1^2-\xi_2^2)\over (p^2-(\xi_1+\xi_2)^2)(p^2-(\xi_1-\xi_2)^2)\,\mathcal F_2(x_1,1,p^2,\underline
    \xi^2)\right)\,.
\end{equation}

These operators can be obtained  from the operator $td/dt+1$ derived
in \S\ref{sec:bubble1} and the change of variables
$t=\frac{\sqrt{\left(p^2-\xi_1^2-\xi_2^2\right)^2-4 \xi_1^2
    \xi_2^2}}{\xi_1^2}$.  For the boundary term one needs to pay
attention that the shift induces a dependence on the physical
parameters in the domain of integration.

%--------------------------------------------------------------------------
\subsubsection{The massive one-loop sunset Feynman integral}
\label{sec:massive-bubble}

Having determined the differential operators acting on the maximal cut
it is now easy to obtain the action of these operators on the one-loop
integral.
The action of the Picard-Fuchs operators on the Feynman
integral $I_\circ(p^2,\xi_1^2,\xi_2^2,2)$ are given by
\begin{equation}\label{e:ode1}
L_{PF,(1)}^{\circ}  \, I_\circ (p^2,\xi_1^2,\xi_2^2,2)  = -{2\over  (p^2-(\xi_1+\xi_2)^2)(p^2-(\xi_1-\xi_2)^2)}\,,
\end{equation}
and
\begin{equation}
 L_{PF,(2)}^\circ \, I_\circ (p^2,\xi_1^2,\xi_2^2,2)= {\xi_1^2-\xi_2^2-p^2\over  (p^2-(\xi_1+\xi_2)^2)(p^2-(\xi_1-\xi_2)^2)}\,.
\end{equation}

It is then easy to obtain that in $D=2$ dimensions the one-loop massive bubble
evaluates to
\begin{multline}
  I_\circ(p^2,\xi_1^2,\xi_2^2) =  {1\over \sqrt{(p^2-(\xi_1+\xi_2)^2)(p^2-(\xi_1-\xi_2)^2)}} \cr
\times \log\left(
    p^2-(\xi_1^2+\xi_2^2)-\sqrt{(p^2-(\xi_1+\xi_2)^2)(p^2-(\xi_1-\xi_2)^2)}\over p^2-(\xi_1^2+\xi_2^2)+\sqrt{(p^2-(\xi_1+\xi_2)^2)(p^2-(\xi_1-\xi_2)^2)}\right) \,.
\end{multline}

%%%%%%%%%%%%%%%%%%%%%%%%%%%%%%%%%%%%%%%%%%%%%%%%%%%%%%%%%%%%%%%%%
\subsection{The two-loop sunset}
\label{sec:two-loop-sunset-1}

The sunset graph  polynomial is the most general cubic in $\mathbb
P^2$  with maximal order two degree for each variables
\begin{equation}
  \mathcal
  F_\su(x_1,x_2,x_3,t,\underline\xi^2)=x_1x_2x_3\left(p^2-(\xi_1^2x_1+\xi_2^2x_2+\xi_3^2x_3)\left({1\over
      x_1}+{1\over x_2}+{1\over x_3}\right)\right)\,,
\end{equation}
which corresponds to the toric polynomial
\begin{equation}
\mathcal F_\su^{tor}=x_1x_2x_3\left(\frac{x_{3} z_{1}}{x_{1}} + \frac{x_{2} z_{2}}{x_{1}} + \frac{x_{3} z_{3}}{x_{2}} + \frac{x_{1} z_{4}}{x_{3}} + \frac{x_{2} z_{5}}{x_{3}} + \frac{x_{1} z_{6}}{x_{2}} + z_{7}\right)\,.
\end{equation}
To the contrary to the one-loop case there are more toric parameters
$z_i$ than physical variables. The identification of the physical
variables is
\begin{equation}\label{e:physpar3}
-\xi_1^2 =z_4=z_6,\quad
-\xi_2^2 =z_2=z_5, \quad
-\xi_3^2 = z_1=z_3,\quad
p^2-(\xi_1^2  +\xi_2^2+ \xi_3^2) =  z_7,
\end{equation}
As before writing the toric polynomial as
\begin{equation}
  P_\su= \sum_{i=1}^7 z_i x_1^{a_i^1} x_2^{a_i^2} x_3^{a_i^3}  \,,
\end{equation}
and setting $\mathbf a_i= (1,a_i^1,a_i^2,a_i^3)$ we have
\begin{equation}
\mathbf  A_\su= \begin{pmatrix}
    \mathbf a_1\cr \vdots\cr \mathbf a_7
  \end{pmatrix}=
  \begin{pmatrix}
    1 & -1 & 0 & 1 \cr
1 & -1 & 1 & 0 \cr
1 & 0 & -1 & 1 \cr
1 & 1 & 0 & -1 \cr
1 & 0 & 1 & -1 \cr
1 & 1 & -1 & 0 \cr
1 & 0 & 0 & 0
  \end{pmatrix}\,,
\end{equation}
The lattice is now defined by
\begin{equation}
\mathbb L_\su:=\{\pmb \ell:=(\ell_1,\dots,\ell_7)\in \mathbb
Z^7|\ell_1\mathbf a_1+\cdots+\ell_7\mathbf a_7=\pmb \ell\cdot \mathbf
A_\su=0 \}\,.
\end{equation}
This lattice in $\mathbb Z^7$ has rank four  $\mathbb
L_\su=\oplus_{i=1}^4  L_i \mathbb Z$ with the basis

\begin{equation}\label{e:BasisKer}
   \begin{pmatrix}
      L_1\cr \vdots\cr  L_4
  \end{pmatrix}=
  \begin{pmatrix}
1 & 0 & 0 & 0 & 1 & 1 & -3 \cr
0 & 1 & 0 & 0 & 0 & 1 & -2 \cr
0 & 0 & 1 & 0 & 1 & 0 & -2 \cr
0 & 0 & 0 & 1 & -1 & -1 & 1
  \end{pmatrix}\,,
\end{equation}
From this we derive the sunset GKZ system
\begin{align}\label{e:GKZsunset1}
 e_1&:= \frac{\partial^{3}}{\partial z_{1}\partial z_{5}\partial z_6} -
  \frac{\partial^{3}}{(\partial z_{7})^{3}},\cr
e_2&:= \frac{\partial^{2}}{\partial z_2\partial z_6} -
  \frac{\partial^{2}}{(\partial z_{7})^{2}},\cr
e_3&:=  \frac{\partial^{2}}{\partial z_{3}\partial z_{5}} -
  \frac{\partial^{2}}{(\partial z_{7})^{2}},\cr
e_4&:= \frac{\partial^{2}}{\partial z_{4}\partial z_{7}} -\frac{\partial^{2}}{\partial z_{5}\partial z_{6}}
\end{align}
by construction $e_i(\mathcal F_\su^{tor})^\alpha=0$ with $\alpha\in\mathbb
C$ for $1\leq i\leq 4$.
We have as well this second set of operators from the operators

\begin{align}\label{e:GKZsunset2}
d_1&:=\sum_{r=1}^7  z_{r} \frac{\partial}{\partial z_{r}},\cr
d_2&:=z_{1} \frac{\partial}{\partial
     z_{1}} + z_{2} \frac{\partial}{\partial z_{2}} -z_{4} \frac{\partial}{\partial z_{4}} - z_{6}
     \frac{\partial}{\partial z_{6}},\cr
d_3&:= z_{2} \frac{\partial}{\partial z_{2}}  - z_{3} \frac{\partial}{\partial
     z_{3}} +z_{5}
     \frac{\partial}{\partial z_{5}}- z_{6} \frac{\partial}{\partial z_{6}}, \cr
d_4&:=z_{1} \frac{\partial}{\partial z_{1}} + z_{3} \frac{\partial}{\partial z_{3}} - z_{4} \frac{\partial}{\partial z_{4}}- z_{5} \frac{\partial}{\partial z_{5}}
\end{align}
The interpretation of these operators is the following
\begin{itemize}
\item The Euler
operator $d_1\mathcal F_{tor}^\alpha= \alpha\,\mathcal F_{tor}^\alpha$ for
$\alpha\in\mathbb C$.
\item To derive the action of these operators on the maximal cut
  period integral
\begin{equation}
  \pi^{tor}_\su(z_1,\dots,z_7)={1\over (2i\pi)^3}\int_\gamma {1\over
    \mathcal F_\su^{tor}}  \prod_{i=1}^3 dx_i\,,
\end{equation}
we remark that if $ \mathcal F_\su^{tor}= x_1x_2x_3 \,  P_\su$ we have
\begin{align}
  d\left({1\over P_\su} {dx_1\over x_1  }\right)&= {-z_1
                                                  x_1/x_2+z_3x_2+z_4
                                                  x_2/x_1-z_6/x_2\over
                                                  P_\su^2}\,
                                                  {dx_1\over
                                                  x_1}\wedge
                                                  {dx_2\over x_2},\cr
d\left({1\over P_\su} {dx_1\over x_1  }\right)&= -{z_1
                                                  x_1/x_2+z_2x_1-z_4
                                                  x_2/x_1-z_5/x_1\over
                                                  P_\su^2}\,
                                                  {dx_1\over
                                                  x_1}\wedge
                                                  {dx_2\over x_2},
\end{align}
therefore  since the cycle $\gamma$ has no boundary
\begin{align}
d_2 \pi^{tor}_\su&= \int_\gamma d\left(  {1\over P_\su} {dx_1\over x_1}\right)=0,\cr
d_3
\pi^{tor}_\su&=-\int_\gamma d\left({1\over P_\su} {dx_2\over x_2}\right)=0, \cr
d_4\pi^{tor}_\su&=\int_\gamma d\left({1\over P_\su} \left({dx_1\over x_1}+{dx_2\over x_2}\right)\right)=0\,.
\end{align}
\item The natural toric coordinates are
\begin{equation}
  u_1:={z_1z_5z_6\over z_7^3},\qquad
u_2:={z_2z_6\over z_7^2},\qquad
u_3:={z_3z_5\over z_7^2},\qquad
u_4:={z_4z_7\over z_5z_6}\,,
\end{equation}
which reads in terms of the physical parameters
\begin{align}
u_2&=\frac{\xi^2_{1} \xi^2_{2}}{{\left( p^2- (\xi^2_{1} + \xi^2_{2} + \xi^2_{3})\right)}^{2}},\quad
u_3=\frac{\xi^2_{2} \xi^2_{3}}{{\left( p^2- (\xi^2_{1} + \xi^2_{2} + \xi^2_{3})\right)}^{2}},\cr
u_4&=\frac{ p^2- (\xi^2_{1} + \xi^2_{2} + \xi^2_{3} )}{\xi^2_{2}},\quad
u_1=u_2u_3u_4.
\end{align}
They are the natural variables associated with the
toric symmetries of the period integral
\begin{align}
    (x_1,x_2)&\to (\lambda x_1,x_2), &(z_1,z_2,z_3,z_4,z_5,z_6,z_7)\to
(z_1/\lambda,z_2/\lambda,z_3,z_4\lambda,z_5\lambda,z_6,z_7),\cr
(x_1,x_2)&\to (x_1,\lambda x_2), &(z_1,z_2,z_3,z_4,z_5,z_6,z_7)\to
(z_1 \lambda,z_2,z_3/\lambda,z_4/\lambda,z_5,z_6\lambda,z_7),\cr
(x_1,x_2)&\to (\lambda  x_1,\lambda x_2), &(z_1,z_2,z_3,z_4,z_5,z_6,z_7)\to
(z_1,z_2/\lambda,z_3/\lambda,z_4,z_5\lambda ,z_6\lambda,z_7).
\end{align}
\end{itemize}

\medskip
The sunset GKZ hypergeometric series is defined as  for
$\gamma_i\not\in\mathbb Z$ with $1\leq i\leq 7$

\begin{equation}
  \Phi^\su_{\mathbb L,\pmb \gamma}(z_1,\dots,z_7)= \sum_{\pmb \ell\in\mathbb
    L} \prod_{i=1}^7 {z_i^{l_i+\gamma_i}\over  \Gamma(l_i+\gamma_i+1)},
\end{equation}
in this sum we have $\pmb \ell= \sum_{i=1}^4 n_i L_i$ with
$n_i\in\mathbb Z$, and
the condition $\sum_{i=1}^7 \gamma_i \mathbf a_i=(-1,0,0,0)$
which can be solved using $\pmb \gamma=(\gamma_1,\dots,\gamma_7)=\sum_{i=1}^4
\gamma_i \mathcal L_i+(0,\dots,0,-1)$. Using the  leading to toric variables
the solution reads
\begin{multline}
  \Phi^\su_{\mathbb L,\pmb \gamma}(z_1,\dots,z_7)= {1\over z_7} \sum_{(n_1,\dots,n_4)\in\mathbb
    Z} {u_1^{n_1+\gamma_1} u_2^{n_2+\gamma_2}
    u_3^{n_3+\gamma_3}u_4^{n_4+\gamma_4}\over
    \prod_{i=1}^4\Gamma(n_i+\gamma_i+1)}\times\cr
\times \, {1\over
   \Gamma(n_1+n_2-n_4+\gamma_1+\gamma_2-\gamma_4+1)\Gamma(n_1+n_3-n_4+\gamma_1+\gamma_3-\gamma_4+1)}\cr
 \times {1\over\Gamma(-3n_1-2n_2-2n_3+n_4-3\gamma_1-2\gamma_2-2\gamma_3+\gamma_4)}\,.
 \end{multline}
With $\pmb \gamma=(0,0,0,0,0,0,0)$ the series is trivially zero
as being resonant. The resolution is to the regularise the term has a
zero by using for $\ell_7<0$
\begin{equation}
  \lim_{\epsilon\to0} {\Gamma(\epsilon)\over \Gamma(\ell_7+\epsilon)}= (-1)^{\ell_7}\Gamma(1-\ell_7)\,,
\end{equation}
and write the associated regulated period as
\begin{multline}
  \pi_\su^{(2)}(p^2,\underline\xi^2) =\lim_{\epsilon\to0} \!\!\!\!\!  \sum_{(n_1,n_2,n_3,n_4)\in \mathbb N}
{} {(\xi_1^2)^{n_1+n_2}(\xi_2^2)^{n_1+n_2+n_3-n_4}(\xi_3^2)^{n_1+n_3}\over\prod_{i=1}^4
    \Gamma(1+n_i)} \cr
\times { (p^2-(\xi_1^2+\xi_2^2+\xi_3^2))^{-3n_1-2n_2-2n_3+n_4-1} (-1)^{-3n_1-2n_2-2n_3+n_4}\Gamma(\epsilon)\over \Gamma(1+n_1+n_2-n_4)\Gamma(1+n_1+n_3-n_4) \Gamma(-3n_1-2n_2-2n_3+n_4+\epsilon)}\,.
\end{multline}
One can expand this expression as a series near $t=\infty$ to get that
\begin{equation}\label{e:pidef2}
  \pi_\su^{(2)}(p^2,\xi_1^2,\xi_2^2,\xi_3^2)= \sum_{n\geq0} (p^2)^{-n-1}\,\sum_{n_1+n_2+n_3=n} \left(n!\over
    n_1!n_2!n_3!\right)^2 \xi_1^{2n_1}\xi_2^{2n_2}\xi_3^{2n_3}\,,
\end{equation}
which is the series expansion of  the maximal cut integral
\begin{equation}
  \pi_\su^{(2)}(p^2,\underline\xi^2)= {1\over (2i\pi)^3}\int_\gamma {1\over
    \mathcal F_\su}\prod_{i=1}^3dx_i\,,
\end{equation}
where $\gamma=\{|x_1|=|x_2|=|x_3|=1\}$.
The construction generalises easily to the case of the  higher loop
sunset integral in an easy way~\cite{DNV}.

\subsubsection{The differential operators}
\label{sec:griff-dwork-meth}

Now that we have the expression for the maximal cut it is easy to
derive the minimal order differential operator annihilating this
period. There are various methods to derive the  Picard-Fuchs operator
from the maximal cut. One method is to use the series expansion of the period around
$s=1/t=0$. Another method is to reduce the GKZ system of differential
operator in similar fashion as shown for the hypergeometric function
in \S\ref{sec:gauss-hyperg-seri}. This method leads to a fourth
order differential operator which factorises a minimal second order
operator. We notice that this approach is similar to the
integration-by-part based approach

The minimal order differential operator is of second order
\begin{equation}
 \mathcal L_{PF}^\su= \left(p^2{d\over dp^2}\right)^2+ q_1(p^2,\underline
 \xi^2)\,
\left( p^2{d\over dp^2}\right)+q_0(p^2,\underline \xi^2)  \,,
\end{equation}
with the coefficients given in\cite{Adams:2013nia,Bloch:2016izu}.
The action of this differential operator on the maximal cut is given
by
\begin{equation}
   \mathcal  L_{PF}^\su\pi_\su^{(2)}= {1\over (2i\pi)^3}\int_\gamma
 \mathcal    L_{PF}^\su{1\over \mathcal F_\su} \,\prod_{i=1}^3dx_i
= {1\over (2i\pi)^3}\int_\gamma
\left(     \sum_{i=1}^3\partial_i\beta_i\right) \,\prod_{i=1}^3dx_i
=0\,.
\end{equation}
The action of this operator on the Feynman integral is given by
then we find that that full differential operator acting on the
two-loop sunset integral is given by

\begin{equation}
  \mathcal   L_{PF}^\su\,I_\su(p^2,\underline \xi^2)= \int_{x_1\geq0\atop x_2\geq0}
\left(     \sum_{i=1}^3\partial_i\beta_i\right)\,\delta(x_3=1) \,\prod_{i=1}^3dx_i
=\mathcal S_\su\,,
\end{equation}
where the inhomogeneous term reads
\begin{equation}\label{e:Ssundef}
\mathcal  S_\su = \mathcal Y_\su(p^2,\underline
  \xi^2)+ c_1(p^2,\underline
  \xi^2) \log\left(m_1^2\over m_3^2\right)+ c_2(p^2,\underline
  \xi^2)
  \log\left(m_2^2\over m_3^2\right)  \,,
\end{equation}
with the Yukawa coupling\footnote{This quantity is the usual Yukawa
  coupling of particle physics and string theory compactification. The
  Yukawa coupling is determined geometrically by the integral of the
  wedge product of differential forms over particular
  cycles~\cite{Candelas:1990rm}. The Yukawa couplings which depend
  non-trivially on the internal geometry appear naturally in the
  differential equations satisfied by the periods of the underlying
  geometry as explained for instance in these reviews~\cite{Morrison:1991cd,Hosono:1994av}.}
\begin{equation}
\mathcal
Y_\su(p^2,\underline
  \xi^2)={6(p^2)^2-4p^2(\xi_1^2+\xi_2^2+\xi_3^2)-2\prod_{i=1}^4\mu_i
\over
(p^2)^2\prod_{ =1}^4 (p^2-\mu_i^2)}\,,
\end{equation}
where
$(\mu_1,\dots,\mu_4)=((-\xi_1+\xi_2+\xi_3)^2), (\xi_1-\xi_2+\xi_3)^2),
(\xi_1+\xi_2-\xi_3)^2), (\xi_1+\xi_2+\xi_3)^2))$.
A  geometric interpretation is the integral~\cite{Bloch:2016izu}
\begin{equation}\label{e:YukawaDef}
 \mathcal  Y_\su(p^2,\underline
  \xi^2) = \int_{\mathcal E_\su}  \Omega_\su\wedge p^2{d\over p^2}\Omega_\su  \,,
\end{equation}
where $\Omega_\su$ is the sunset residue differential form
\begin{equation}
  \Omega_\su=\textrm{Res}_{\mathcal E_\su=0} {x_1dx_2\wedge dx_3+x_3dx_1\wedge dx_2+x_2 dx_3\wedge
    dx_1\over \mathcal F_\su}\,,
\end{equation}
on the sunset elliptic curve
\begin{equation}\label{e:EsunsetDef}
\mathcal E_\su:=\{ p^2 x_1x_2x_3 -
(\xi_1^2x_1+\xi_2^2x_2+\xi_3^2x_3)(x_1x_2+x_1x_3+x_2x_3)| (x_1,x_2,x_3)\in\IP^2\}\,.
\end{equation}
The Yukawa coupling satisfies the differential equation
\begin{equation}
 p^2{d\over p^2}  \mathcal Y_\su(t)= (2-q_1(p^2,\underline\xi^2) )\mathcal Y_\su(p^2,\underline\xi^2)\,.
\end{equation}
The coefficients $c_1$ and $c_2$  in~\eqref{e:Ssundef}  are the integral of the residue one
form between the marked points on $Q_1=[0,-\xi_3^2,\xi_2^2]$,
$Q_2=[-\xi_3^2,0,\xi_1^2]$ and $Q_3=[-\xi_2^2,\xi_1^2,0]$ on the
elliptic curve~\cite{Bloch:2016izu}
\begin{equation}
  c_1(p^2,\underline
  \xi^2):= p^2{d\over p^2}\int_{Q_1}^{Q_3} \Omega_\su, \qquad  c_2(p^2,\underline
  \xi^2):= p^2{d\over p^2}\int_{Q_2}^{Q_3} \Omega_\su\,.
\end{equation}

%-------------------------------------------------------------------------
\subsection{The generic case}
\label{sec:generic-case}

In this section we show how to determine the differential equation for
the $l$-loop sunset integral from the knowledge of the maximal cut.
The maximal cut of the $l$-loop sunset integral is given by
\begin{equation}\label{e:Pisudef}
\pi_\su^{(l)}(p^2,\underline \xi^2)= \sum_{n\geq0} t^{-n-1} \,  A_\su(l,n,\xi_1^2,\dots,\xi_{l+1}^2)\,,
\end{equation}
with
\begin{equation}\label{e:AperyDef}
  A_\su(l,n,\xi_1^2,\cdots,\xi_{l+1}^2):=
\sum_{r_1+\cdots+r_{l+1}=n} \left(n!\over r_1!\cdots r_{l+1}!\right)^2
  \prod_{i=1}^{l+1} \xi_i^{2r_i}\,.
\end{equation}

\subsubsection{The all equal mass case}
\label{sec:all-equal-mass}

For the all equal mass case one can easily determine the differential
equation to all order~\cite{Vanhove:2014wqa} using the Bessel integral
representation of~\cite{Bailey:2008ib}. We present here a
different derivation.

\medskip
For the all equal masses the coefficient of the maximal cut satisfies a nice
recursion~\cite{Verril2004}

\begin{equation}
  \sum_{k\geq0} \left(n^{l+2} \sum_{1\leq i\leq
      k}\sum_{a_i+b_i=l+2\atop 1<a_{i+1}+1<a_i\leq l+1} \prod_{i=1}^k
      (-a_ib_i) \left(n-i\over n-i+1\right)^{a_i-1}\right) \,
    A_\su(l,n-k,\underline 1)=0  ,
\end{equation}
where $a_i\in \mathbb N$.
Standard method gives that the associated differential operator acting on
$t\pi^l_\su(t,1,\dots,1)=\sum_{n\geq0} (p^2)^{-n} A(l+1,n,1,\dots,1)$ reads
\begin{multline}
  \mathcal L_{PF,\su}^{(l),1mass}= \sum_{k\geq0} (p^2)^{k} \sum_{1\leq i\leq
    k}\sum_{a_i+b_i=l+2, a_{k+1}=0\atop 1<a_{i+1}+1<a_i\leq l+1}
  \left(k-  p^2{d\over p^2}\right)^{l+2-a_1}\cr
  \times\prod_{i=1}^k (-a_ib_i) \left(  k-i-p^2{d\over dp^2}\right)^{a_i-a_{i+1}}.
\end{multline}
This operator has been derived in~\cite[\S9]{Vanhove:2014wqa} using different method.

They are differential operators of  order  $l$,
the loop order, in $d/dp^2$ and  the  coefficients are polynomials of   degree $l+1$
\begin{equation}
  \mathcal L_{PF}^{(l),1mass}=(-p^2)^{\lceil l/2\rceil-1}\prod_{i=1}^{\lfloor l/2\rfloor+1} (p^2-\mu_i^2)    \left({d\over dp^2}\right)^l  + \cdots
\end{equation}
where $\mu_i^2:=(\pm 1\pm 1\cdots \pm1)^2$  is the set of the different
thresholds.
The operator $ \mathcal L_{PF}^{(2),1mass}$ is the Picard-Fuchs operator of the
family of elliptic curves for $\Gamma_1(6)$ for the
all equal mass sunset~\cite{Bloch:2013tra},  the operator  $ \mathcal L_{PF}^{(3),1mass}$  of the
family of $K3$ surfaces~\cite{Bloch:2014qca}.
Having determined the Picard-Fuchs operator it is not difficult to
derive its action on the Feynman integral with the result that~\cite{Vanhove:2014wqa}

\begin{equation}
  \mathcal   L_{PF}^{(l),1mass} ( I_\su(p^2,1,\dots,1))= -(l+1)!\,.
\end{equation}

\subsubsection{The general mass case}
\label{sec:general-mass-case}

For unequal masses the recursion relation does not close only on the
coefficients~\eqref{e:AperyDef} and no simple closed formula is known
for the differential operator on the maximal cut.
The minimal differential operator annihilating the $\pi_\su^{(l)}
(t,\underline \xi^2)$ can be obtained using  the GKZ
hypergeometric function discussed in  the previous section.

For the $l$-loop sunset integral the GKZ lattice has rank $l^2$,
$\mathbb L=\sum_{i=1}^{l^2}n_i \, L_i$.
For instance for the three-loop sunset
the regulated hypergeometric series representation of the maximal cut reads
\begin{multline}
\pi_\su^{(3)}(p^2,\underline\xi^2) =-\lim_{\epsilon\to0}\sum_{(n_1,\dots,n_9)\in\mathbb N^9} { (\xi_1^2)^{n_1+n_2+n_3}
  (\xi_2^2)^{n_1+n_3+n_4+n_6-n_7-n_8+n_9}
\over  \prod_{i=1}^9 \Gamma(1+n_i)  } \cr
\times {  (\xi_3^2)^{n_2+n_5+n_8} (\xi_4^2)^{n_1+n_4+n_6}\over\Gamma
   (n_1+n_4+n_6-n_7-n_8+1) \Gamma
   (n_2+n_5-n_6+n_8-n_9+1) }\cr
 \times{
  \over \Gamma
   (n_1+n_3-n_5+n_6-n_7-n_8+n_9+1)}\cr
\times{
   (-p^2+\xi_1^2+\xi_2^2+\xi_3^2+\xi_4^2)^{-3
   n_1-2 n_2-2 n_3-2 n_4-n_5-2n_6+n_7-n_9-1} \Gamma (\epsilon )\over\Gamma (-3
   n_1-2 n_2-2 n_3-2 n_4-n_5-2
   n_6+n_7-n_9+\epsilon )}.
\end{multline}
The  minimal order differential operator annihilating the maximal cut
$ p^2\pi_\su^{(3)}(p^2,\underline\xi^2)$
 with
generic mass configurations, $\xi_1\neq\xi_2\neq\xi_3\neq\xi_4$ and
all the masses non vanishing, is an operator of order 6, with
polynomial coefficients $c_k(t)$ of degree up to 29
\begin{equation}
  L_{PF,\su}^3=\sum_{k=0}^6 c_k(t) \left(t{d\over t}\right)^k \,.
\end{equation}
For instance the differential
operator for the mass configuration $\xi_i=i$ with $1\leq i\leq 4$ is
given by
\begin{align}
  c_6&= (t-100)  (t-36)  (t-64)  (t-4)^2  (t-16)^2\cr
&\times\big( 345  t^{12}-10275  t^{11}+243243  t^{10}+700860  t^9-289019444  t^8+9517886160  t^7\cr
&-169244843904  t^6+2163112875520  t^5-24375264125952  t^4\cr
&+198627459010560  t^3-896517312217088  t^2\cr
&+1570362910310400  t-1192050032640000\big),
\end{align}
and
\begin{align}
c_5&= (t-4)  (t-16)
\big(7245  t^{17}-1461150  t^{16}+108842709  t^{15}-4073021820
     t^{14}\cr
&+79037467036  t^{13}+706049613520  t^{12}-122977114948800  t^{11}\cr
&+4897976525794560  t^{10}
-118057966435402752  t^9\cr
&+2042520337021317120  t^8-28129034886941589504  t^7\cr
&+321784682881513881600  t^6-2877522528057659228160  t^5\cr
&+17978948962533528043520  t^4
-69950845277551433089024  t^3\cr
&+151178557780128065126400  t^2
-182250696371318292480000  t\cr
&+96676211287130112000000\big),
\end{align}
and
\begin{align}
c_4&=2\, \big(
    23460  t^{19}-4086975  t^{18}+273974766  t^{17}-9833465295  t^{16}\cr
&+173874227860  t^{15}
+3780156754180  t^{14}\cr
&-419091386081744  t^{13}+16647873781420800  t^{12}\cr
&-425729411677916160  t^{11}+8098824799795968000  t^{10}\cr
  &-125136842089603031040  t^9
    +1631034274362173030400  t^8\cr
&+17364390414642101354496  t^7+
140612615518097533829120  t^6 \cr
&-807868060015143792148480  t^5+3100095209313936311582720  t^4\cr
&-7563751451192001262780416  t^3 +
11448586013594218187980800  t^2\cr
&-9812428506034109153280000  t+3374878648568905728000000     \big),
\end{align}
and
\begin{align}
c_3&=12\big(
   8970  t^{19}-1147050  t^{18}+56442264  t^{17}-1477273050  t^{16}-447578647  t^{15}\cr
&+2416587481200  t^{14}-130189239609348  t^{13}+4001396495500560  t^{12}\cr
&-86975712270293184  t^{11}+1511724058206439680  t^{10}\cr
&-22690173944998831104  t^9
+289974679497600921600  t^8\cr
&-2900762618196498137088  t^7
+20882244400635484241920  t^6 \cr
&-101090327023260610854912  t^5
+308760428925736546467840  t^4\cr
&-559057237244267332632576  t^3
+533177283118109609164800  t^2\cr
&-133034777312420167680000  t
-140619943690371072000000\big),
\end{align}
and
\begin{align}
 c_2&=24  \big(3105  t^{19}-260100  t^{18}+8740695  t^{17}-121279200  t^{16}-8982728081  t^{15}\cr
&+771645247175  t^{14}
-29786960482306  t^{13}+741851366254700  t^{12}\cr
&-14140682364004072  t^{11}
+237224880534337760  t^{10}
\cr
&-3605462277123620992  t^9
+44725169880349560320  t^8
\cr
&-405767142088142927872  t^7 
+2549108215435181793280  t^6 \cr 
&-11307241496864563101696  t^5
+40972781273200446013440  t^4 \cr
&-141797614014479525216256  t^3
+363118631232748702924800  t^2 \cr
&-415180490608717332480000  t
+210929915535556608000000 \big), \cr 
\end{align}
and
\begin{align}
c_1&=24\,\big( 345  t^{19}-15000  t^{18}+345675  t^{17}+7323600  t^{16}-3165461083  t^{15}\cr
&+184943420750  t^{14}
-5084383561348  t^{13}+91042473303800  t^{12}\cr
&-1344824163401536
t^{11}+17444484465759680  t^{10}\cr
&-146155444722244096  t^9 -426434786380119040  t^8\cr
&+31798683088486989824  t^7
-488483076656283893760  t^6\cr
&+5136134162164414021632  t^5 -40834519838668015534080  t^4 \cr
&+222597043391679285952512  t^3-685074395310881085849600  t^2 \cr
&+830360981217434664960000  t-421859831071113216000000 \big),
\end{align}
and
\begin{align}
c_0&=1728\,\big( 21908444  t^{15}-1482071825  t^{14}+40507170144  t^{13}-668436089250  t^{12}\cr
&+8209054542408  t^{11}-65000176183240  t^{10}-503218239747392  t^9\cr
&+31962708303867520  t^8
-619576476284137472  t^7+7554395788685281280  t^6\cr
&-73455221906789646336  t^5
+571135922816871792640  t^4\cr
&-3095113137012548304896  t^3
+9514922157095570636800  t^2 \cr
&
-11532791405797703680000  t+5859164320432128000000 \big)\,.
\end{align}
A systematic study of the differential operators for the $l$ loop
sunset integral will appear in~\cite{DNV}.

%%%%%%%%%%%%%%%%%%%%%%%%%%%%%%%%%%%%%%%%%%%%%%%%%%%%%%%%%%%%%%%%%%
\section{Analytic evaluations for  sunset integral}\label{sec:sunset}

In this section we give different  analytic expressions for the
two-loop sunset integral. In one form the two-loop sunset integral is
given by an elliptic dilogarithm as review in
\S\ref{sec:sunsetdilog} or as a ordinary trilogarithm as review
in \S\ref{sec:sunsetdilog}.  In \S\ref{sec:mirror} we
explain that the equivalence between the two expressions is a
manifestation  of the mirror symmetry proven in~\cite{Bloch:2016izu}.

%-------------------------------------------------------------------------
\subsection{The sunset integral as an elliptic dilogarithm}
\label{sec:sunsetdilog}

The geometry of the graph hypersurface is a family of elliptic
curves
\begin{equation}
\mathcal E_\su:=\{ p^2 x_1x_2x_3 -
(\xi_1^2x_1+\xi_2^2x_2+\xi_3^2x_3)(x_1x_2+x_1x_3+x_2x_3)| (x_1,x_2,x_3)\in\IP^2\}\,.
\end{equation}
One can use the information from the geometry of the graph polynomial
and use a parameterisation of the physical variables making the
geometry of the elliptic curve explicit.

The elliptic curve $\mathcal E_\su$ can be represented as $\mathbb
C^\times/q^{\mathbb Z}$ where $q=\exp(2i\pi\tau)$ and $\tau$ is the
period ratio of the elliptic curve.
There a six special points on the elliptic curve $\mathcal E_\su$ the
three points that intersect the domain of integration
\begin{equation}
  P_1:=[1,0,0], \qquad P_2:=[0,1,0],\qquad P_3:=[0,0,1],
\end{equation}
and three other points outside the domain of integration
\begin{equation}
  Q_1:=[0,-\xi_3^2,\xi_2^2], \qquad Q_2:=[-\xi_3^2,0,\xi_1^2],\qquad Q_3:=[-\xi_1^2,\xi_2^2,0].
\end{equation}
If one denotes by $x(P_i)$ the image of the point $P_i$ in  $\mathbb
C^\times/q^{\mathbb Z}$ and $x(Q_i)$ the  image of the point $Q_i$ we
have $ x(P_i)=-x(Q_i)$ with $i=1,2,3$
\begin{equation}
\left(\theta_1(x(P_i)/x(P_j))\over \theta_c(x(P_i)/x(P_j))\right)^2=
            {\xi_k\over \sqrt t\xi_i\xi_j}\,,
\end{equation}
with $(i,j,k)$ a permutation of $(1,2,3)$ and $c$ a permutation of
$(2,3,4)$.\footnote{The
Jacobi theta functions are defined by
$\theta_2(q):=2q^{1\over8}\prod_{n\geq1} (1-q^n)(1+q^n)^2$,
$\theta_3(q):= \prod_{n\geq1}(1-q^n)(1+q^{n-\frac12})^2$ and
$\theta_4(q):=\prod_{n\geq1}(1-q^n)(1-q^{n-\frac12})^2$.
}
It was  shown in~\cite{Bloch:2016izu} that the sunset Feynman integral is given by
\begin{equation}\label{e:I1}
 I_\su(p^2,\underline\xi^2)\equiv {i \varpi_r   \over\pi} \left(\hat E_2\left(x(P_1)\over
      x(P_2)\right)+\hat E_2\left(x(P_2)\over x(P_3)\right)+\hat
    E_2\left(x(P_3)\over x(P_1)\right)\right)\,
\mod  \textrm{periods}\, ,
\end{equation}
where $\hat E_2(x)$ is the elliptic dilogarithm
\begin{equation}\label{e:E2h}
  \hat E_2(x)= \sum_{n\geq0} \left(\Li_2(q^nx)-\Li_2(-q^nx)\right)
    -\sum_{n\geq1} \left(\Li_2(q^n/x)-\Li_2(-q^n/x)\right)  \, .
\end{equation}

\medskip

The $J$-invariant of the sunset elliptic curve is
\begin{equation}
  J_\su  = 256{(3-u_\su^2)^3\over 4-u_\su^2}\,,
\end{equation}
where the Hauptmodul is
\begin{equation}
  u_\su={(p^2-\xi_1^2-\xi_2^2-\xi_3^2)^2-4(\xi_1^2\xi_2^2+\xi_1^2\xi_3^2+\xi_2^2\xi_3^2)\over \sqrt{16t\xi_1^2\xi_2^2\xi_3^2}}  ,
\end{equation}
  given in term of Jacobi theta functions
\begin{equation}
u_\su^{3,4}={\theta_3^4+\theta_4^4\over \theta_3^2\theta_4^2}, \quad
 u_\su^{2,3}=-{\theta_3^4+\theta_2^4\over \theta_3^2\theta_2^2}, \quad
u_\su^{2,4}=i{\theta_2^4-\theta_4^4\over \theta_2^2\theta_4^2},
\end{equation}
and the period is given for each pair $(a,b)=(3,4), (2,3), (2,4)$ by
\begin{equation}
\varpi_r={t^{1\over4} \pi \theta_a\theta_b\over (\xi_1^2\xi_2^2\xi_3^2)^{1\over4}},
\end{equation}
 is the elliptic curve  period which is  real
on the line $t<(\xi_1+\xi_2+\xi_3)^{2}$.

By using the dilogarithm functional equations one can bring the
expression~\eqref{e:I1} in a form similar to the one used in~\cite{Brown:6917B}
\begin{equation}
\sum_{i=1}^3  \sum_{n\in\mathbb Z} \textrm{Li}_2(q^n x_i)  \,.
\end{equation}
This representation needs to be properly regularised as discussed
in~\cite{Brown:6917B} whereas the representation in~\eqref{e:E2h} is
a converging sum.  An equivalent representation used multiple elliptic
polylogarithms~\cite{Adams:2014vja,Adams:2015gva,Adams:2015ydq,Adams:2016vdo,Adams:2018ulb,Adams:2018yqc}
this representation  has the advantage of generalising to other graphs~\cite{Broedel:2017kkb,Broedel:2017siw,Broedel:2018iwv,Broedel:2018rwm,Broedel:2018tgw,Remiddi:2017har}.

\bigskip
For the all equal masses case, $1=\xi_1=\xi_2=\xi_3$, the family of elliptic curves
\begin{equation}\label{e:Es1mass}
\cEs:=\{p^2 x_1x_2x_3- (x_1+x_2+x_3)(x_1x_2+x_1x_3+x_2x_3)=0|(x_1,x_2,x_3)\in\mathbb P^2\},
\end{equation}
defines a
pencil of elliptic curves in $\mathbb P^2$ corresponding to a modular
family of elliptic curves $f : \cE_\su \to X_1(6) = \{\tau \in \mathbb C|
\Imm(\tau) > 0\}/\Gamma_1(6)$ (see~\cite{Bloch:2013tra}).
When all the masses are equal the map is easier
since the elliptic curve is a modular curve for $\Gamma_1(6)$ and the
coordinates of the points are mapped to sixth  root of unity
$x(P_r)=e^{2i\pi r\over 6}$ and $x(Q_r)=-e^{2i\pi r\over 6}$  with $r=1,2,3$.

The integral is expressed as the following combination of elliptic
dilogarithms
\begin{equation}\label{e:Intsunset}
 I_\su(p^2,1,1,1)=\varpi_r(t) (i\pi-\log q)
-6 {\varpi_r(p^2)\over\pi} \,E_\circleddash(q) \,,
\end{equation}
where the Hauptmodul
\begin{equation}\label{e:thaupt}
p^2=9+72{\eta(q^2)\over \eta(q^3)}\left(\eta(q^6)\over\eta(q)\right)^5\,,
\end{equation}
and the real period for $p^2<\xi_1^2+\xi_2^2+\xi_3^2$
\begin{equation}
\varpi_r(p^2)={\pi\over\sqrt3}\,
   {\eta(q)^6\eta(q^6)\over \eta(q^2)^{3}\eta(q^3)^{2}}\,.
\end{equation}
 In this case the   elliptic dilogarithm  is given by
\begin{eqnarray}\label{e:Esunset}
    E_\circleddash(q) &=&- {1\over2i} \sum_{n\geq0}  \left(\Li2(q^n\zeta_6^5)
    +\Li2(q^n\zeta_6^4)-\Li2(q^n\zeta_6^2) -\Li2(q^n\zeta_6)\right)\cr
&+&{1\over 4i} \, \left(\Li2(\zeta_6^5)
    +\Li2(\zeta_6^4)-\Li2(\zeta_6^2) -\Li2(\zeta_6)\right)\,.
\end{eqnarray}
which we can write as a $q$-expansion
\begin{equation}
  E_\circleddash(q)={1\over2}\,\sum_{k\in \mathbb Z\backslash \{0\}}{(-1)^{k-1}\over k^2}\, {\sin({n\pi\over3})+\sin({2n\pi\over3})\over
  1-q^k}\,.
\end{equation}

\subsection{The sunset integral as a trilogarithm}
\label{sec:sunsettrilog}

In this section we evaluate the sunset two-loop integral in a
different way, leading to an expression in terms of trilogarithms. We
leave the interpretation of the two equivalence with the previous
evaluation to \S\ref{sec:mirror} where we explain that these
results are a  manifestation of local mirror symmetry.

We introduce the quantity the logarithmic Mahler measure $R_0(p^2,\underline\xi^2)$
\begin{equation}\label{e:R0def}
 R_0(p^2,\underline xi^2)=- i\pi+ \int_{|x|=|y|=1}\hspace{-.6cm}
  \log(p^2-(\xi_1^2x+\xi_2^2y+\xi_3^2)(x^{-1}+y^{-1}+1)) \,{d\log x
    d\log y\over(2\pi i)^2}\,,
\end{equation}
which evaluates to
\begin{equation}
  R_0= \log(-p^2) -\sum_{n\geq1}  {(p^2)^{-n}\over n} A_\su(2,n,\xi_1^2,\xi_2^2,\xi_3^2)\,,
\end{equation}
where $A_\su(2,n,\xi_1^2,\xi_2^2,\xi_3^2)$ is defined in~\eqref{e:AperyDef}.
Differentiating with respect to $p^2$ leads to maximal cut
\begin{equation}
  {d\over dp^2} R_0(p^2,\xi_1^2,\xi_2^2,\xi_3^2)= \pi^{(2)}_\su(p^2,\xi_1^2,\xi_2^2,\xi_3^2),
\end{equation}
where $\pi^{(2)}_\su(p^2,\xi_1^2,\xi_2^2,\xi_3^2)$ is defined in~\eqref{e:Pisudef}.
It was shown in~\cite{Bloch:2016izu} that the sunset integral has the
expansion
\begin{equation}\label{e:curiousInt}
  I_\su(p^2,\underline \xi^2)=-2i\pi\,\pi_\su^{(2)}(t,\underline\xi^2)\,\left( 3 R_0^3  +\!\!\!\!\!\!\!\!\!\!\!\!\sum_{\ell_1+\ell_2+\ell_3=\ell>0\atop (\ell_1,\ell_2,\ell_3)\in\IN^3\backslash(0,0,0)}\!\!\!\!\!\!\!\!\!\!\!\! \!\!\!\!\!\!\ell (1- \ell
  R_0)
  N_{\ell_1,\ell_2,\ell_3}\,\prod_{i=1}^3  \xi_i^{2\ell_i}e^{\ell_iR_0}\right),
\end{equation}
where the invariant numbers  $N_{\ell_1,\ell_2,\ell_3}$ can be
computed from the Yukawa coupling~\eqref{e:YukawaDef} using~\cite[proposition~7.6]{Bloch:2016izu}
\begin{equation}
  6- \sum_{\ell_1+\ell_2+\ell_3=\ell>0\atop (\ell_1,\ell_2,\ell_3)\in\IN^3\backslash(0,0,0)}  \ell^3 N_{\ell_1,\ell_2,\ell_3} R_0^\ell\,\prod_{i=1}^3\xi_i^{2\ell_i}=
  {(6(p^2)^2-4p^2(\xi_1^2+\xi_2^2+\xi_3^2)+2\mu_1\cdots\mu_4)\over
   p^2\prod_{i=1}^4(p^2-\mu_i^2) \,(\pi_\su^{(2)}(p^2,\underline\xi^2))^3 } .
\end{equation}
These quantities can be
expressed in terms of the virtual integer numbers of
rational curves  of degree $\ell=\ell_1+\ell_2+\ell_3$ by the covering formula
\begin{equation}\label{e:Nton}
  N_{\ell_{1},\ell_{2},\ell_{3}}=\sum_{d|\ell_{1},\ell_{2},\ell_{3}}\frac{1}{d^{3}}n_{\frac{\ell_{1}}{d},\frac{\ell_{2}}{d},\frac{\ell_{3}}{d}}\,.
\end{equation}
A first few Gromov-Witten numbers are given by (these invariants are
symmetric in their indices so list only one representative)
\begin{equation}\label{e:GWtable}
\begin{tabular}{|c||c|c|c|c|c|c|c|c|c|}
\hline
$(\ell_1,\ell_2,\ell_3)$ & $(100)$ & $\overset{k>0}{(k00)}$ & $(110)$ & $(210)$ & $(111)$ & $(310)$ & $(220)$ & $(211)$ & $(221)$\tabularnewline
\hline
\hline
$N_{\ell_{1},\ell_{2},\ell_{3}}$ & $2$ & $2/k^{3}$ & $-2$ & $0$ & $6$ & $0$ & $-1/4$ & $-4$ & $10$\tabularnewline
\hline
$n_{\ell_{1},\ell_{2},\ell_{3}}$ & $2$ & $0$ & $-2$ & $0$ & $6$ & $0$ & 0 & $-4$ & $10$\tabularnewline
\hline
\end{tabular}
\end{equation}
\begin{equation}
\begin{tabular}{|c||c|c|c|c|c|c|c|c|c|}
\hline
$(\ell_1,\ell_2,\ell_3)$ & $(410)$ & $(320)$ & $(311)$ & $(510)$ & $(420)$ & $(411)$ & $(330)$ & $(321)$ & $(222)$\tabularnewline
\hline
\hline
$N_{\ell_{1},\ell_{2},\ell_{3}}$ & $0$ & $0$ & $0$ & $0$ & $0$ & $0$ & $-2/27$ & $-1$ & $-189/4$\tabularnewline
\hline
$n_{\ell_{1},\ell_{2},\ell_{3}}$ & $0$ & $0$ & $0$ & $0$ & $0$ & $0$ & $0$ & $-1$ & $-48$\tabularnewline
\hline
\end{tabular}
\end{equation}
Introducing the variables $Q_i=\xi_i^2 e^{R_0}$ we can rewrite the
sunset integral as
 \begin{multline}\label{e:3triexp}
  -{I_\su( p^2 ,\underline\xi^2)\over 2i\pi \pi_\su^{(2)}( p^2 ,\underline \xi^2)}= 3R_0^3+   \sum_{\ell_1+\ell_2+\ell_3=\ell>0\atop (\ell_1,\ell_2,\ell_3)\in\IN^3\backslash(0,0,0)}\ell (1- \ell
   R_0)    N_{\ell_1,\ell_2,\ell_3}\,\prod_{i=1}^3
   \xi_i^{2\ell_i}e^{\ell_iR_0}\cr
=   3R_0^3+\sum_{(n_1,n_2,n_3)\geq(0,0,0)}
   \,(d_{n_1,n_2,n_3}+\delta_{n_1,n_2,n_3}\,\log(-p^2))\,
   \textrm{Li}_3(Q_1^{n_1}Q_2^{n_2}Q_3^{n_3})\,,
\end{multline}
where $\textrm{Li}_3=\sum_{n\geq1} x^n/n^3$ is the trilogarithm and
the first coefficients are given by
\begin{eqnarray}\label{e:ddtilde}
\begin{tabular}{|c||c|c|c|c|c|c|c|c|c|c|}
\hline
$(\ell_1,\ell_2,\ell_3)$ & $(100)$ & $(110)$&$(200)$ & $(111)$ &
                                                                 $(210)$ &$(300)$ &$(400)$& $(220)$ & $(310)$ & $(211)$ \tabularnewline
\hline
\hline
$d_{\ell_{1},\ell_{2},\ell_{3}}$ & $2$ & $0$ & $9/4$ & $-6$&$-6$ & $58/27$&$79/48$ & $0$ & $-8/3$ & $40$\tabularnewline
\hline
$\delta_{\ell_{1},\ell_{2},\ell_{3}}$ & $-2$&$2$ & $8$ & $-54$ & $0$ & $-16/27$ & $-3/8$ & $3$ & $0$ & $64$\tabularnewline
\hline
\end{tabular}\nonumber\\
\end{eqnarray}
In \S\ref{sec:mirror} we will explain
that these numbers are local
Gromov-Witten numbers $N_{\ell_1,\ell_2,\ell_3}$ and the sunset
Feynman integral is the Legendre transformation of the local prepotential as shown~\cite{Bloch:2016izu}.
\medskip

Using the relation between the complex structure of $2i\pi\tau=\log q$ of the
elliptic curve and $R_0$ (see~\cite[proposition~7.6]{Bloch:2016izu} and \S\ref{sec:mirror})

\begin{equation}\label{e:mirrormap}
  \log q=   2\sum_{i=1}^3\log(Q_i^2)
-\!\!\! \!\!\!\!\sum_{\ell_1+\ell_2+\ell_3=\ell>0\atop (\ell_1,\ell_2,\ell_3)\in\IN^3\backslash(0,0,0)} \!\!\!\! \!\!\!\!
  \ell^2 N_{\ell_1,\ell_2,\ell_3} \, \prod_{i=1}^3 Q_i^{\ell_i},
\end{equation}
one can check the equivalence between the expressions~\eqref{e:I1} and~\eqref{e:curiousInt}.

%-------------------------------------------------------------------------
\subsubsection{The all equal masses case}
\label{sec:grom-witt-invar-1}

In this section we compute the local invariants for the all equal
masses case $\xi_1=\xi_2=\xi_3=1$ the sunset integral reads
\begin{equation}\label{e:IsuGW1mass}
 I_\su(p^2, 1,1,1)=\pi_\su^{(2)}(p^2, 1,1,1)\,\left( 3 R_0^3  +\!\!\!\!\! \!\!\!\!\!\sum_{\ell_1+\ell_2+\ell_3=\ell>0\atop (\ell_1,\ell_2,\ell_3)\in\IN^3\backslash(0,0,0)}\!\!\!\!\! \!\!\!\!\! \!\!\!\!\ell (1- \ell
  \log Q)
  N_{\ell_1,\ell_2,\ell_3}\ Q_0^{\ell}\right)\,.
\end{equation}
with $Q_0= \exp(R_0)$ where
\begin{equation}
  R_0=-\log(-p^2)+  \sum_{\ell>0} {(p^2)^{-\ell}\over\ell}  \sum_{p_1+p_2+p_3=\ell} \left(\ell!\over p_1!p_2!p_3!\right)^2\,,
\end{equation}
and using the expression for $p^2$ in~\eqref{e:thaupt}  we have that
\begin{equation}
 R_0(q)= i\pi+ \log q   -\sum_{n\geq1} (-1)^{n-1} \left(-3\over n\right)\,n\,\Li_1(q^n)\,,
\end{equation}
where $\left(-3\over n\right)=0,1,-1$ for $n\equiv 0,1,2\mod 3$. The
maximal cut in~\eqref{e:pidef2}
reads
\begin{equation}
  p^2\pi_\su^{(2)}(p^2,1,1,1) = {\eta(q^2)^6\eta(q^3)\over \eta(q)^3\eta(q^6)^2}\,.
\end{equation}
We recall the $p^2$ is the hauptmodul in~\eqref{e:thaupt}.
The Gromov-Witten invariant $N_\ell$ can be computed using~\cite[proposition~7.6]{Bloch:2016izu}
\begin{equation}\label{e:Nk}
  6-\sum_{\ell\geq1} \ell^3 N_\ell Q^\ell=  {6\over  p^2 (p^2-1 )( p^2-9 )\, (\pi_\su^{(2)}(q))^3}\,.
\end{equation}
Introducing  the virtual numbers $n_{\ell}$ of degree $\ell$
\begin{equation}
   N_\ell = \sum_{d|\ell} {1\over d^3} n_{\ell\over d},
\end{equation}
we have
\begin{align}
	n_k/6 &=1,-1,1,-2,5,-14,42,-136,465,-1655,6083,-22988,\\
	\notag &\quad\ 88907,-350637,1406365, -5724384,
   23603157,-98440995,\\
	\notag &\quad\ 414771045,-1763651230,7561361577,-32661478080,\\
	\notag &\quad\ 14204649044
   1,-621629198960,2736004885450,\\
	\notag &\quad\ -12105740577346,53824690388016,\dots
\end{align}
The relation between $Q$  and $q$
\begin{equation}\label{e:Qtoq}
 Q=-q\, \prod_{ n\geq1}  (1-q^n)^{n \delta(n)}; \qquad \delta(n):=(-1)^{n-1} \,
\left(-3\over n \right)\,,
\end{equation}
which we will interpret as a mirror
map in \S\ref{sec:mirror}, in the expansion
in~\eqref{e:IsuGW1mass} gives the  dilogarithm expression in~\eqref{e:Intsunset}.

\subsection{Mirror symmetry and sunset integral}\label{sec:mirror}

In this section we review the result of~\cite{Bloch:2016izu} where it
was shown that the sunset two-loop integral is the Legendre transform of the local
Gromov-Witten prepotential and that the equivalence between the
elliptic dilogarithm expression and the trilogarithm expansion of the previous
section is a manifestation of local mirror symmetry. The techniques used in this section are standard in the study of mirror symmetry in string theory. We refer to the physicists oriented reviews~\cite{Hosono:1994av,Closset:2009sv} for some presentation of the mathematical notions used in this section. 

\subsubsection{The sunset graph polynomial and del Pezzo surface}
\label{sec:suns-graph-polyn}

\begin{figure}[h]
\label{fig:hexagon}
  \centering
\includegraphics[width=6cm]{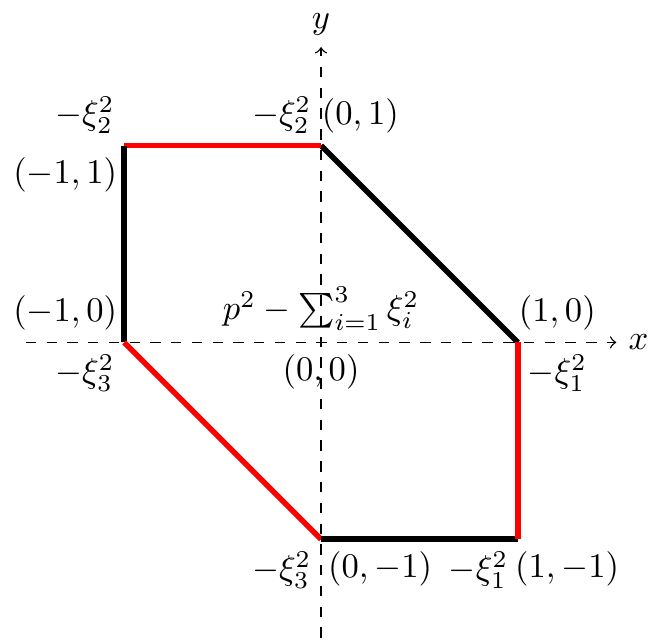}
\caption{The Newton polyhedron associated with the sunset second
  Symanzik polynomial. The coordinates $(a,b)$ of the vertices we the powers
  of $x^a y^b$ and we give the value of the coefficient in $\phi_\su(p^2,\underline\xi^2,x,y,1)$.}
\end{figure}

To the sunset Laurent polynomial
\begin{equation}
P_\su(p^2,\underline\xi^2,x_1,x_2,x_3)=p^2-(x_1\xi_1^2+x_2\xi_2^2+x_3\xi_3^2)\left({1\over
    x_1}+{1\over x_2}+{1\over x_3}\right),
\end{equation}
we  associate the Newton polyhedron in figure~\ref{fig:hexagon}.
The vertices of the polyhedron are the powers of the monomial in $x_1$
and $x_2$ with $x_3=1$.

This corresponds to a maximal toric blow-up of three points in
$\mathbb P^2$ leading to a del Pezzo surface of degree 6 $\mathcal
B_3$.\footnote{A  del Pezzo surface  is a two-dimensional Fano variety.  A Fano variety is a complete variety whose anti-canonical bundle  is ample. The anti-canonical bundle of a non-singular algebraic variety  of dimension $n$  is the line bundle  defined as the $n$th exterior power of the inverse of the cotangent bundle. An ample line bundle is a bundle with enough global sections to set up an embedding of its base variety or manifold  into projective space.  } The hexagon in figure~\ref{fig:hexagon} resulted from the blow-up (in
red on the figure) of a
triangle at the points $P_1=[1:0:0]$, $P_2=[0:1:0]$ and $P_3=[0:0:1]$
by the mass parameters see~\cite[\S6]{Bloch:2013tra} and~\cite[\S4]{Bloch:2016izu}.   The del Pezzo 6
surfaces are rigid.\footnote{
The graph polynomial~\eqref{e:Fsunsetdef} for higher loop sunset graphs defines Fano
variety, which is as well a Calabi-Yau manifold.}

Notice that the external momentum $p^2$ appears only in the centre of
the Newton polytope making this variable special.

One can construct a non-compact Calabi-Yau three-fold $\mathcal M_\su$ defined as the
anti-canonical hypersurface over the del Pezzo surface  $\mathcal B_3$.   This non-compact three-fold is
obtained as follows~\cite[\S5]{Bloch:2016izu}.
Consider the Laurent polynomial
\begin{equation}\label{e:FsuDef}
  F_\su= a+ b u^2v^{-1} +c u^{-1}v+ u^{-1}v^{-1}  \phi_\su(p^2,\underline\xi^2,x_1,x_2,x_3),
\end{equation}
with $a,b,c\in\mathbb C^*$.  Its Newton polytope  $\Delta$ is the convex hull
of $\{(0,0,2,-1),  \break (0,0,-1,1), \Delta_\su\times (-1,-1)\}$  where
$\Delta_\su$ is the Newton polytope given by the hexagon in
figure~\ref{fig:hexagon}. The newton polytope $\Delta$ is reflexive
because its polar polytope
$\Delta^\circ:=\{y\in\mathbb R^4|\langle y,x\rangle\geq-1,\forall x
\in\Delta\}=\textrm{convex~hull}\{(0,0,1,0), (0,0,0,1), 6\Delta^\circ_\su\times(-2,-3)\}$
is integral. Notice that for the sunset polytope is self-dual
$\Delta_\su=\Delta^\circ_\su$.
A triangulation of $\Delta$ gives a complete toric fan\footnote{The
  fan of a toric variety is defined in the standard reference~\cite{Fulton} and the review oriented to a physicts audience in~\cite{Closset:2009sv}.} on $\Delta^\circ$, which  then
provides  Fano variety $\mathbb P_{\Delta}$ of
dimension four~\cite{Coxtoric}. For general $a,b,c$ and the generic physical
parameters $p^2,\xi_1^2,\xi_2^2,\xi_3^2$ in the sunset graph
polynomial, the singular compactification
$\mathcal M_\su:= \overline{\{  F=0\}}$
is a smooth Calabi-Yau three-fold.
This non-compact Calabi-Yau three-fold can be seen as a limit of
compact Calabi-Yau three-fold following the
approach of~\cite{CKYZ} to local mirror symmetry.
One can consider a  semi-stably degenerating a family of
elliptically-fibered Calabi-Yau three-folds $\mathcal M_{z}$
to a singular compactification $\mathcal M_\su$ for $z=0$
and to compare the asymptotic Hodge theory\footnote{Feynman integrals are  period integrals of mixed Hodge structures~\cite{Bloch:2005bh,Vanhove:2014wqa}. At a singular point  some cycles of integration vanish, the so-called vanishing cycles, and the limiting behaviour of the period integral is captured by the asymptotic behaviour of the cohomological Hodge theory. The asymptotic Hodge theory inherit some filtration and weight structure of the original Hodge theory.} of this B-model to that of
the mirror (elliptically fibered) A-model Calabi-Yau
$\mathcal M_\su^{\circ}$. Both $\mathcal M_\su$ and
$\mathcal M_\su^{\circ}$ are elliptically fibered  over the del Pezzo
of degree 6 $\mathcal B_3$.  Under the mirror map we have the isomorphism of A- and B-model $\mathbb{Z}$-variation of Hodge structure~\cite{Bloch:2016izu}
\begin{equation}
 H^3(\mathcal M_{z_0}) \cong
 H^{even}(\mathcal M^{\circ}_{q_0}) \,.
\end{equation}

This situation is not unique to the two-loop sunset. The sunset graph
have a reflexive polytopes containing the origin. The origin of the
polytope is associated with the coefficient $p^2-\sum_{i=1}^n
\xi_i^2$, and plays a very special  role. The ambient space of  the sunset polytope
defines a Calabi-Yau hypersurfaces (the anti-canonical divisor defines
a Gorenstein toric Fano variety).  Therefore they are a natural home
for Batyrev's mirror symmetry
techniques~\cite{Batyrev94}.

\subsubsection{Local mirror symmetry}
\label{sec:local-mirr-symm}

Putting this into practise means recasting the computation in \S\ref{sec:sunsettrilog}
and the mirror symmetry description in~\cite[\S7]{Bloch:2016izu} in
the language of~\cite{Huang:2013yta}, matching the computation of the
Gromov-Witten prepotential in~\cite[\S6.6]{Huang:2013yta}.

The first step is to remark that the holomorphic $(3,0)$ period of
Calabi-Yau three-fold $\mathcal M_\su$ reduces to the third period
$R_0$ once integrated on a vanishing
cycle~\cite[Appendix~A]{Hosono:2006ga},~\cite[\S4]{Katz:1996fh} and~\cite[\S5.7]{Bloch:2016izu}
\begin{equation}
  \int_{\rm vanishing~cycle} \textrm{Res}_{F_\su=0} \left({1\over F_\su} {du\wedge
  dv\wedge \wedge dx_1\wedge dx_2\over u v x_1x_2}\right)\propto R_0  (p^2,\underline\xi^2),
\end{equation}
where $F_\su$ is given in~\eqref{e:FsuDef} and $R_0  (p^2,\underline\xi^2)$ is given in~\eqref{e:R0def}.
This second period is  related to the
analytic period near $p^2=\infty$ by
$\pi_\su^{(2)}  (p^2,\underline\xi^2)={d\over dp^2}
R_0(p^2,\underline\xi^2)$.\footnote{ It has been already noticed in\cite{Stienstra:2005} the special role played by the Mahler measure and mirror symmetry.}

The Gromov-Witten invariant evaluated in~\eqref{e:GWtable}
section~\ref{sec:sunsettrilog} are actually the BPS numbers for the del
Pezzo 6 case evaluated in~\cite[\S6.6]{Huang:2013yta}  since

\begin{equation}\label{e:Flocal}
\sum_{\ell_1+\ell_2+\ell_3=\ell>0\atop
  (\ell_1,\ell_2,\ell_3)\in\IN^3\backslash(0,0,0)}
  N_{\ell_1,\ell_2,\ell_3}
  R_0^\ell\,\prod_{i=1}^3\xi_i^{2\ell_i}=\!\!\!\!\! \!\!\!\!\!\sum_{(\tilde\ell_1,\tilde\ell_2,\tilde\ell_3)
  \in\IN^3\backslash(0,0,0)}\!\!\!\!\! \!\!\!\!\!
  n_{\tilde\ell_1,\tilde\ell_2,\tilde\ell_3}
  \textrm{Li}_3(\prod_{i=1}^3\xi_i^{2\tilde\ell_i}
  e^{\tilde\ell_iR_0}),
\end{equation}
where we used the covering relation~\eqref{e:Nton}.
 With the following identifications\footnote{We would like
  to thank Albrecht Klemm for discussions and communication that
  helped clarifying the link between the work in~\cite{Bloch:2016izu}
  and the analysis in~\cite{Huang:2013yta}.}
$Q_1=1$, $Q_2=\xi_1^2e^{R_0}$, $Q_3=\xi_2^2 e^{R_0}$ and
$Q_4=\xi_3^2e^{R_0}$, the expression in~\eqref{e:Flocal} reproduces
the  local genus 0  prepotential $F_0=F_0^{\textrm class}+\sum_{\pmb\beta\in
    H^2(\mathcal M,\mathbb Z)} n^{\pmb\beta}_g  \textrm{Li}_3(\prod_{r=1}^4  Q_r^{\beta_r
  }) $ computed in~\cite[eq.(6.51)]{Huang:2013yta} with
  $F_0^{\textrm class}= \prod_{i=1}^3 (R_0+\log(\xi_i^2))$ in our case.

  \medskip
From the complex structure of the elliptic
curve we define the dual period $\pi_1  (p^2,\underline\xi^2)=2i\pi\tau\pi_\su^{(2)}
(p^2,\underline\xi^2)$ one the other homology cycle. Which gives the
dual third period $R_1$,
such that $\pi_1^{(2)}  (p^2,\underline\xi^2)={d\over dp^2}
R_1(p^2,\underline\xi^2)$.
This dual period $R_1$ is therefore identified with  the derivative of local prepotential $F_0$
\begin{align}
  2i\pi R_1&=  {\partial\over \partial R_0} \,F_0\\
\nonumber  &= \sum_{1\leq i<j\leq 3} (R_0+\log(\xi_i^2))
  (R_0+\log(\xi_j^2))
  -\!\!\!\!\! \!\!\!\sum_{\ell_1+\ell_2+\ell_3=\ell>0\atop(\ell_1,\ell_2,\ell_3)\in
    \mathbb N^3\backslash\{0,0,0\}} \!\!\! \!\!\! \!\!\!\ell N_{\ell_1,\ell_2,\ell_3}
  \prod_{i=1}^3 \xi_i^{2\ell_i}e^{\ell_i R_0}\,,
\end{align}
as shown in~\cite[theorem~6.1]{Bloch:2016izu}
and~\cite[Corollary~6.3]{Bloch:2016izu}.
With these identifications it is not difficult to see that the
sunset Feynman integral is actually given by the Legendre   transform of $R_1$
\begin{equation}
  I_\su (p^2,\underline \xi^2)=-{2i\pi} \pi_\su^{(2)}( p^2,\underline\xi^2)
  \left( {\partial R_1\over\partial R_0} R_0- R_1\right)\,.
\end{equation}
This shows the relation between the sunset Feynman integral computes the local
Gromov-Witten prepotential. The local mirror symmetry  map
$Q\leftrightarrow q$ given in
the  relations~\eqref{e:mirrormap} and~\eqref{e:Qtoq} maps the B-model
expression, where the sunset Feynman integral is a  elliptic
dilogarithm function of the complex structure $\log(q)/(2i\pi)$ of the elliptic curve
and the A-model expansion in terms of the K\"ahler moduli $Q_i$.

\section{Conclusion}
\label{sec:conclusion}

In this text we have reviewed the toric approach to the evaluation of the
maximal cut of Feynman integrals and the derivation of the minimal
order differential
operator acting on the Feynman integral.
On the particular example of the sunset integral we have shown that
the Feynman integral can take two different but equivalent forms.
One form is an elliptic polylogarithm but it can
as well expressed as standard trilogarithm. We have explained that
mirror symmetry can be used to evaluate around the point where
$p^2=\infty$. The expressions there makes explicit all the mass
parameters.  One remarkable fact is that the computation can be done
using the existing technology of mirror symmetry developed in other
physical~\cite{Hosono:1993qy,CKYZ,Huang:2013yta} or
mathematics~\cite{Doran:2011ti} contexts. This analysis extends
naturally to the higher loop sunset integrals~\cite{DNV}.
The elliptic polylogarithm representation generalises to other
two-loop integrals like the kite
integral~\cite{Adams:2016xah,Bogner:2017vim,Bogner:2018uus} or the
all equal masses three-loop sunset~\cite{Bloch:2014qca}. This representation leads to fast numerical
evaluation~\cite{Bogner:2017vim}.   But it has the disadvantage of hiding all the physical
parameters in the geometry of the elliptic curve.
The expression using the trilogarithm has the advantage of making all
the mass parameters explicit and generalising to all loop orders since
the expansion of the higher-loop sunset graphs around $p^2=\infty$
is expected to involve polylogarithms of order $l$
at $l$-loop order~\cite{Doran:2011ti,DNV}.

 %------------------------------------------------------------------------
 \section*{Acknowledgements}
It is a pleasure to thank Charles Doran and Albrecht Klemm for discussions.
The research of
P. Vanhove has received funding the ANR grant ``Amplitudes'' ANR-17-
CE31-0001-01, and is partially supported by Laboratory of Mirror
Symmetry NRU HSE, RF Government grant, ag. N$^\circ$ 14.641.31.0001.

%%%%%%%%%%%%%%%%%%%%%%%%%%%%%%%%%%%%%%%%%%%%%%%%%%%%%%%%%%%%%%%%%

\end{document}